\documentclass[manuscript]{aastex}
\usepackage{natbib}
\usepackage{lineno}
\usepackage[figuresright]{rotating}
\usepackage{lscape}
\usepackage{longtable}

\shorttitle{Evolving spectral properties of outbursts  of 4U~1730--22}
\shortauthors{Chen et al.}

\begin{document}


\title{Return of 4U~1730--22 after 49 years silence: 
the outburst properties observed by NICER and  Insight-HXMT
}

\author{Yu-Peng Chen\textsuperscript{*}}
\email{chenyp@ihep.ac.cn}
\affil{Key Laboratory for Particle Astrophysics, Institute of High Energy Physics, Chinese Academy of Sciences, 19B Yuquan Road, Beijing 100049, China}

\author{Shu Zhang\textsuperscript{*}}
\email{szhang@ihep.ac.cn}
\affil{Key Laboratory for Particle Astrophysics, Institute of High Energy Physics, Chinese Academy of Sciences, 19B Yuquan Road, Beijing 100049, China}

\author{Shuang-Nan Zhang\textsuperscript{*}}
\email{zhangsn@ihep.ac.cn}
\affil{Key Laboratory for Particle Astrophysics, Institute of High Energy Physics, Chinese Academy of Sciences, 19B Yuquan Road, Beijing 100049, China}
\affil{University of Chinese Academy of Sciences, Chinese Academy of Sciences, Beijing 100049, China}

\author{Long Ji}
\affil{School of Physics and Astronomy, Sun Yat-Sen University, Zhuhai, 519082, China}

\author{Peng-Ju Wang}
\affil{Key Laboratory for Particle Astrophysics, Institute of High Energy Physics, Chinese Academy of Sciences, 19B Yuquan Road, Beijing 100049, China}
\affil{University of Chinese Academy of Sciences, Chinese Academy of Sciences, Beijing 100049, China}

\author{Ling-Da Kong}
 \affil{ Institut f\"{u}r Astronomie und Astrophysik, Kepler Center for Astro and Particle Physics, Eberhard Karls Universit\"{a}t, Sand 1, D-72076 T\"{u}bingen, Germany}

\author{Zhi Chang}
\affil{Key Laboratory for Particle Astrophysics, Institute of High Energy Physics, Chinese Academy of Sciences, 19B Yuquan Road, Beijing 100049, China}

\author{Jing-Qiang Peng}
\affil{Key Laboratory for Particle Astrophysics, Institute of High Energy Physics, Chinese Academy of Sciences, 19B Yuquan Road, Beijing 100049, China}
\affil{University of Chinese Academy of Sciences, Chinese Academy of Sciences, Beijing 100049, China}

\author{Qing-Cang Shui}
\affil{Key Laboratory for Particle Astrophysics, Institute of High Energy Physics, Chinese Academy of Sciences, 19B Yuquan Road, Beijing 100049, China}
\affil{University of Chinese Academy of Sciences, Chinese Academy of Sciences, Beijing 100049, China}

\author{Jian Li}
\affil{CAS Key Laboratory for Research in Galaxies and Cosmology, Department of Astronomy, University of Science and Technology of China, Hefei 230026, China}
\affil{School of Astronomy and Space Science, University of Science and Technology of China, Hefei 230026, China}

\author{Lian Tao}
\affil{Key Laboratory for Particle Astrophysics, Institute of High Energy Physics, Chinese Academy of Sciences, 19B Yuquan Road, Beijing 100049, China}

\author{Ming-Yu Ge}
\affil{Key Laboratory for Particle Astrophysics, Institute of High Energy Physics, Chinese Academy of Sciences, 19B Yuquan Road, Beijing 100049, China}

\author{Jin-Lu Qu}
\affil{Key Laboratory for Particle Astrophysics, Institute of High Energy Physics, Chinese Academy of Sciences, 19B Yuquan Road, Beijing 100049, China}

\begin{abstract}
After in quiescence for 49 years, 4U~1730--22 became active and had two outbursts in 2021 \& 2022, the onset and tail of the outbursts were observed by NICER, which give us a peerless opportunity to study the state transition and its underlying mechanism.
In this work, we take both the NS surface and accretion disk emission as the seed photons of the Comptonization and derive their spectral evolution in a bolometric luminosity range of 1\%--15\%$L_{\rm Edd}$.
In the high/soft state, the inferred inner disk radius and the NS radius are consistent well, which implies that the accretion disk is close to the NS surface.
For the decay stage, we report a steep change of the accretion disk emission within one day, i.e., the soft-to-hard transition, which could be due to the propeller effect and the corresponding neutron star surface magnetic field is 1.8--2.2$\times10^{8}$ G.
Moreover, the inner disk radius is truncated at the corotation radius, which is similar to the propeller effect detected from 4U~1608--52.
The absence of the propeller effect in the hard-to-soft state transition implies that the transition between the magnetospheric accretion and the disk accretion is not the sole cause of the state transitions.
\end{abstract}

\section{Introduction}

About one third of low mass X-ray binaries (LMXBs) hosting neutron stars (NS) are transient sources and characterized by a luminosity that varies over $\sim10^{32}-10^{38}~{\rm erg}^{-2}{\rm s}^{-1}$, which is likely caused by an instability of the accretion induced by a variable accretion rate, $\dot{M}$. In this picture, the accretion flow drained from the companion star propagates towards the NS from the outer regions of the accretion disk (see review, e.g., \citealp{Done2007}), which is confirmed by multi-wavelength observations (e.g., \citealp{Lopez-Navas2020}).
Except for the accretion rate, the NS's parameters, e.g., magnetic field and spin period, could also affect the accretion process.
In theory, the luminosity's abrupt drop at $\sim10^{36}-10^{37}~{\rm erg}^{-2}~{\rm s}^{-1}$ in NS-LMXBs is produced by a transition from the disk accretion (material drops onto the NS surface) to the magnetospheric accretion (material drops on the magnetosphere and is then expelled out from the system), i.e., the propeller effect \citep{Zhang1998}. {\bf The interaction between the magnetic field and the accretion disk occurs in many types of stars involving young stellar objects, white dwarfs, and NSs, which has influence on the inner disk radius, outflow rate and etc (e.g., \citealp{2006ApJ...646..304U, 2015SSRv..191..339R, 2009A&A...493..809B,2018A&A...610A..46C}).}

  In this picture, the propeller effect happens when the accretion rate is low enough that the magnetospheric radius, $r_{\rm m}$ (at which the ram pressure of the infalling material is balanced against the magnetic pressure) is larger than the corotation radius $r_{\rm c}$ (at which gravitational forces and centrifugal forces are balanced) \citep{Illarionov1975,Cui1997},
\begin{eqnarray}
  r_{\rm c}=1.7\times 10^{8}P_{\rm spin}^{2/3}M_{1.4}^{1/3}~{\rm cm},
   \label{corotation}
\end{eqnarray}
here $P$ is the NS spin period in units of second, $M_{1.4}$ is the NS mass in units of 1.4$M_{\odot}$.
That is, if $r_{\rm m}>r_{\rm c}$ the accretion disk is truncated at $r_{\rm m}$ and the accreted material is mostly ``propelled" out by the spinning magnetosphere, rather than falling down to the NS surface.
The transition causes a luminosity gap \citep{Corbet1996,Campana2000}, $\Delta\equiv r_{\rm m}/r_{\rm NS} \ge r_{\rm c}/r_{\rm NS}$. This gap has a lower limit value, 	$r_{\rm c}/r_{\rm NS}$, which is only dependent on the NS mass $M_{\rm NS}$, NS radius $r_{\rm NS}$, and NS spin $P_{\rm spin}$.

4U~1730--22 was detected by Uhuru in 1972 \citep{Cominsky1978,Forman1978}, and after half a century of quiescence it
returned to be active in 2021 and 2022.
The bright luminosity in the quiescent state \citep{Tomsick2007},
thermonuclear bursts (type I X-ray bursts) \citep{Chen2022b,Li2022}, and the burst oscillation around  $\nu$=584.65 Hz \citep{Li2022}, indicate its NS nature and its magnetic field $\sim~10^{8}-10^{9}$ G. 
NICER and Insight-HXMT made high cadence observations on the two outbursts and covered the whole outburst stage: the onset, the peak, and the extinction, which is an ideal sample to study the accretion process, e.g., the `propeller effect' and the luminosity gap during the state transition.
For 4U~1730--22, $r_{\rm c}$=24 km, and it is predicted  $\Delta=2.4$ by assuming $M_{\rm NS}=$1.4 $M_{\odot}$, $r_{\rm NS}$=10 km.

In this work, using the broad energy band capabilities of Insight-HXMT and the large effective area in soft X-ray band of NICER, we study the two outbursts from 4U~1730--22. 
In previous works, when the outburst spectrum is fitted, the accretion disk emission is usually taken as seed photon component and the NS surface emission is ignored, or the NS surface emission is taken as the seed photon component but the accretion disk emission is ignored, e.g., \citep{Zhang1998,Chen2006}. This is due to that the faint NS surface emission of less than one-tenth of the disk emission when the source is bright or that it is hard to distinguish them due to the constraints of the previous instruments. However, when the accretion rate is very low, i.e., $\sim$ 1\%$L_{\rm Edd}$,
the NS surface emission could not be ignored, and should be involved in the spectral fitting. 
Especially, in the onset and the extinction of the outburst, the NS surface is the main seed photon component and is much brighter than the disk emission. 
In this work, by stacking observations when the source is very faint, 
the disk emission and the NS surface emission are both involved in the spectral fitting, 
and the disk radius and temperature are obtained (Section 2 and Section 3).
Moreover, a state transition with a luminosity's abrupt drop is observed by NICER, which could be caused by the propeller effect (Section 4). 

\section{Observations and Data Reduction}

\subsection{Insight-HXMT}
Insight-HXMT was launched on the 15th of June 2017, which excels in its broad energy band (1--250 keV), large effective area in the hard X-rays energy band and little pile-up for bright sources (up to several Crab) \citep{Zhang2020}.
There are three main payloads, and all of them are collimated telescopes: the High Energy X-ray Telescope (HE; poshwich NaI/CsI, 20--250 keV, $\sim$ 5000 cm$^2$), the Medium Energy X-ray Telescope (ME; Si pin detector, 5--40 keV, 952 cm$^2$) and the Low Energy X-ray telescope (LE; SCD detector, 1--12 keV, 384 cm$^2$). 
For the three main payloads of Insight-HXMT, each has two main field of views (FoVs), i.e., LE: 1.6$^{\circ}\times6^{\circ}$ and 6$^{\circ}\times6^{\circ}$, ME: 1$^{\circ}\times4^{\circ}$ and 4$^{\circ}\times4^{\circ}$, HE: 5.7$^{\circ}\times1.1^{\circ}$ and 5.7$^{\circ}\times5.7^{\circ}$.
Moreover,they also have the blind FoV (full blocked) detectors, which are used for the background estimation and energy calibration.

\begin{figure}[t]
 \centering
   \includegraphics[angle=90, scale=0.6]{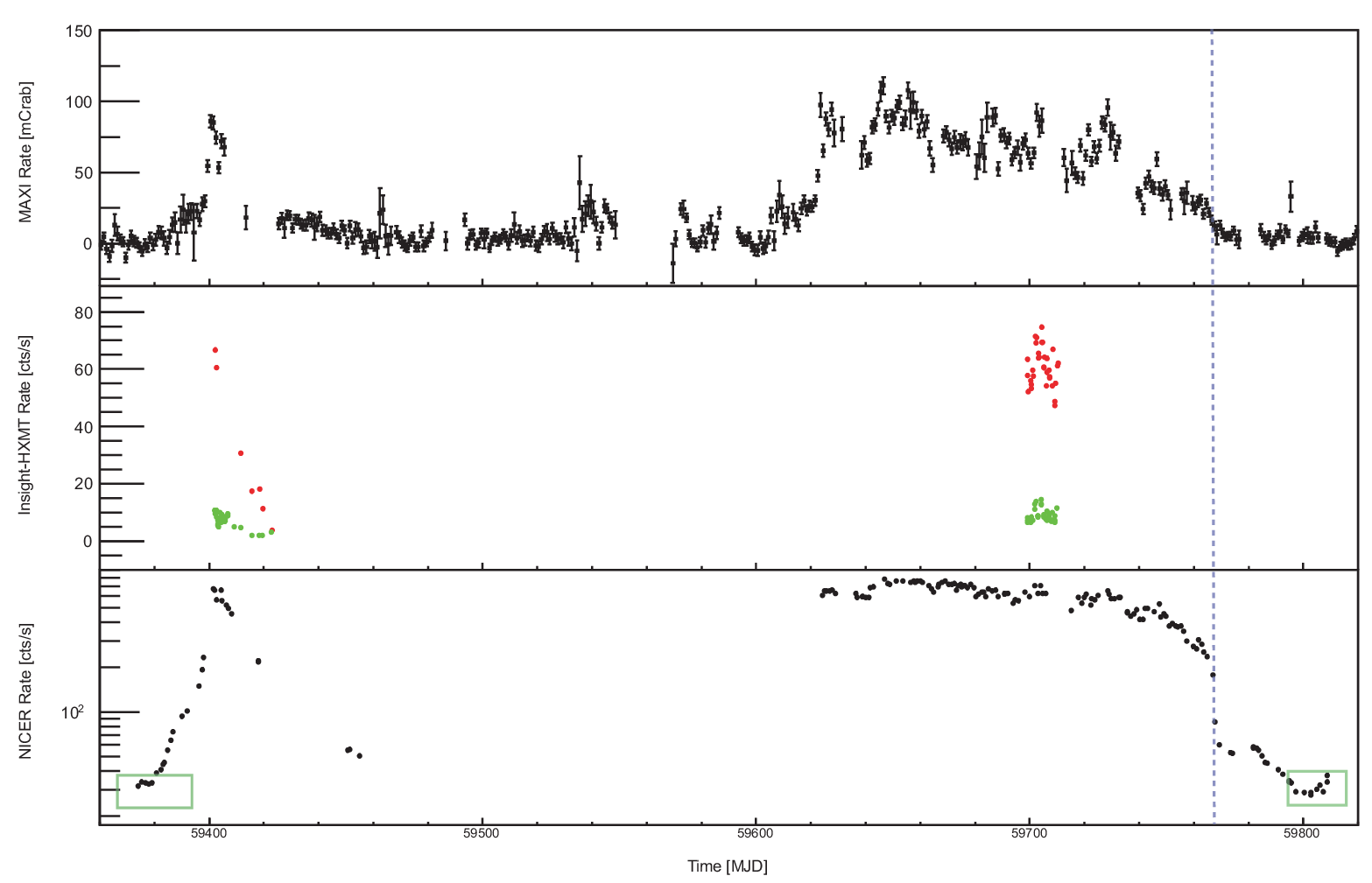}
 \caption{Top panel: daily light curves of 4U~1730--22 by MAXI (black, 2--20 keV)   during the  outburst  in  2021 and 2022. 
Middel panel: Net light curves of 4U~1730--22  by LE (1--10 keV, red) and ME (8--30 keV, green) which are rebinned by one obsid ($\sim$ 10000 s).  
Bottom panel: Net light curves of 4U~1730--22 by NICER (0.3--12 keV, its full energy band)  which are rebinned by one obsid ($\sim$ 3000 s). Please note the error bars of the light curves of LE, ME, and NICER are smaller than the size of the symbols; and the NICER lightcurves are plotted in  a diagram of log count rate versus time.
The 'propeller effect' time is marked by a blue line.
{\bf The data points in the green box indicate the obsids with count rate $<$ 40 cts/s at the onset of the outburst in 2021 and the extinction of the outburst in 2022, which are stacked and fitted respectively; the derived model parameters from the two stacked specta are shown in Table \ref{tb_fit_thcomp_bb}. }
 }
 \label{fig_lc_nicer_le_me}
\end{figure}

As shown in Figure \ref{fig_lc_nicer_le_me}, for the two outbursts in 2021 and 2022, Insight-HXMT observed 4U~1730--22 with 74 observations ranging from P041401100101-20210707-01-01 to P051400201402-20220513-02-01 with a total observation time of 184 ks.
These observations covered the peak and decay phases of the outburst in 2021 and the peak stage of the outburst in 2022.
The HE spectrum is not involved in the joint spectral fitting of the persistent emission, since the expected HE flux falls below the systematic error of the background model. 


We use the Insight-HXMT Data Analysis software (HXMTDAS) v2.05\footnote{http://hxmtweb.ihep.ac.cn/} to extract the lightcurves, the spectra and background emission following the recommended procedure of the Insight-HXMT Data Reduction. 
For the spectral fitting of LE and ME, the energy bands are chosen to be 2--7 keV and 8--20 keV. 
The spectra of LE and ME are rebinned by ftool ftgrouppha optimal binning algorithm \citep{Kaastra2016} with a minimum of 25 counts per grouped bin, and a 1\% systematic error is added to account for the uncertainties of the background model and calibration \citep{Li2020}.
The resulting spectra are analyzed using XSPEC \citep{Arnaud1996} version 12.12.0.

\subsection{NICER}

For the two outbursts, NICER also performed high cadence observations, which covered almost all the stages of the two outbursts but missed the onset and rise stage of the outburst in 2022.
NICER has observed 4U~1730--22 with 162 observations ranging from 4202200101 to 4639010210 with a total exposure time of 372 ks.

The NICER data are reduced using the pipeline tool nicerl2\footnote{https://heasarc.gsfc.nasa.gov/docs/nicer/nicer\_analysis.html} in NICERDAS v7a with the standard NICER filtering and using ftool XSELECT to extract lightcurves and spectra.
Among the 52 operational detectors, the Focal Plane Module (FPM) No. 14 and 34 are removed from the analysis because of increased detector noise.
The response matrix files (RMFs) and ancillary response files (ARFs) are generated with the ftool nicerrmf and nicerarf.
The background is estimated using the ftool nibackgen3C50 \citep{Remillard2022}. 
As we did for the Insight-HXMT spectra, the spectra of NICER are rebinned by ftool ftgrouppha optimal binning algorithm \citep{Kaastra2016} with a minimum of 25 counts per grouped bin.

The tbabs model with Wilm abundances accounts for the ISM absorption in the spectral model \citep{Wilms2000}.
We added a systematic uncertainty of 1\% to the NICER spectrum in 1--10 keV and 5\% to the NICER spectrum in 0.4--1 keV because of a larger systematic uncertainty caused by the NICER instrument and the unmodelled background.

A color-color diagram (CCD) is attempted to plot from NICER data or MAXI data, however, the trajectory in the plot is not a coherent way to represent systems, which could be due to the narrow energy band of NICER and low sensitivity of MAXI. Thus we plot a hardness-intensity diagram (HID), as shown in Figure \ref{fig_hid}, where `hardness' is defined as the count rate ratio (2.5--10 keV/1.7--2.5 keV) and `intensity' is defined the count rate in 0.4--10 keV. Each point on HID represents an obsid with an exposure time of $\sim$1000--5000 s. The black points and red points represent the outbursts in 2021 and 2022, respectively.

\section{Analysis and Results}

\begin{figure}[t]
 \centering
   \includegraphics[angle=0, scale=0.5]{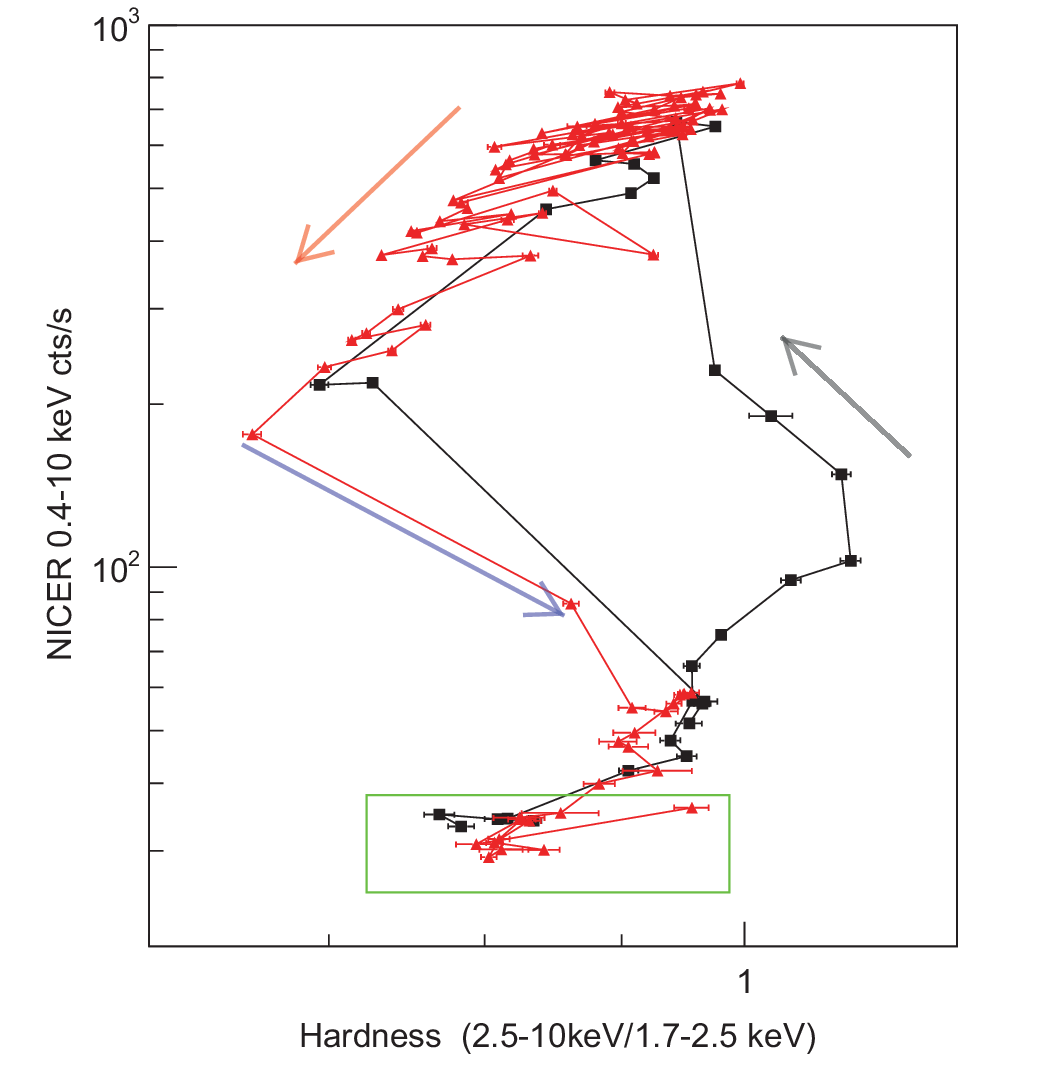}
 \caption{Hardness-intensity diagram for NICER observations. Hardness is defined as the NICER count-rate ratio  in the (2.5--10 keV)/(1.7--2.5 keV) bands, intensity is defined as the count rate of NICER in 0.4--10 keV. Each point in the diagram represents one obsid with an exposure time of  $\sim$1000-5000 s. The black points and red points represent the outburst in 2021 and 2022, respectively. For the outburst in 2021, it began at the   bottom left and evolved in an anticlockwise loop, {\bf as shown in the gray arrow}. For the outburst in 2022, the NICER observations missed the outburst's onset and the first observation is located at the top right and evolved in an anticlockwise  direction, ending at the bottom left, {\bf as shown in the red arrow}. The 'propeller effect' time is marked by a blue arrow. 
 {\bf The black and red points in the green box indicate the onset of the outburst in 2021 and the extinction of the outburst in 2022, which are $<$ 40 cts/s and stacked and fitted respectively.} 
 }
\label{fig_hid}
\end{figure}

\begin{figure}[t]
 \centering
   \includegraphics[angle=0, scale=0.6]{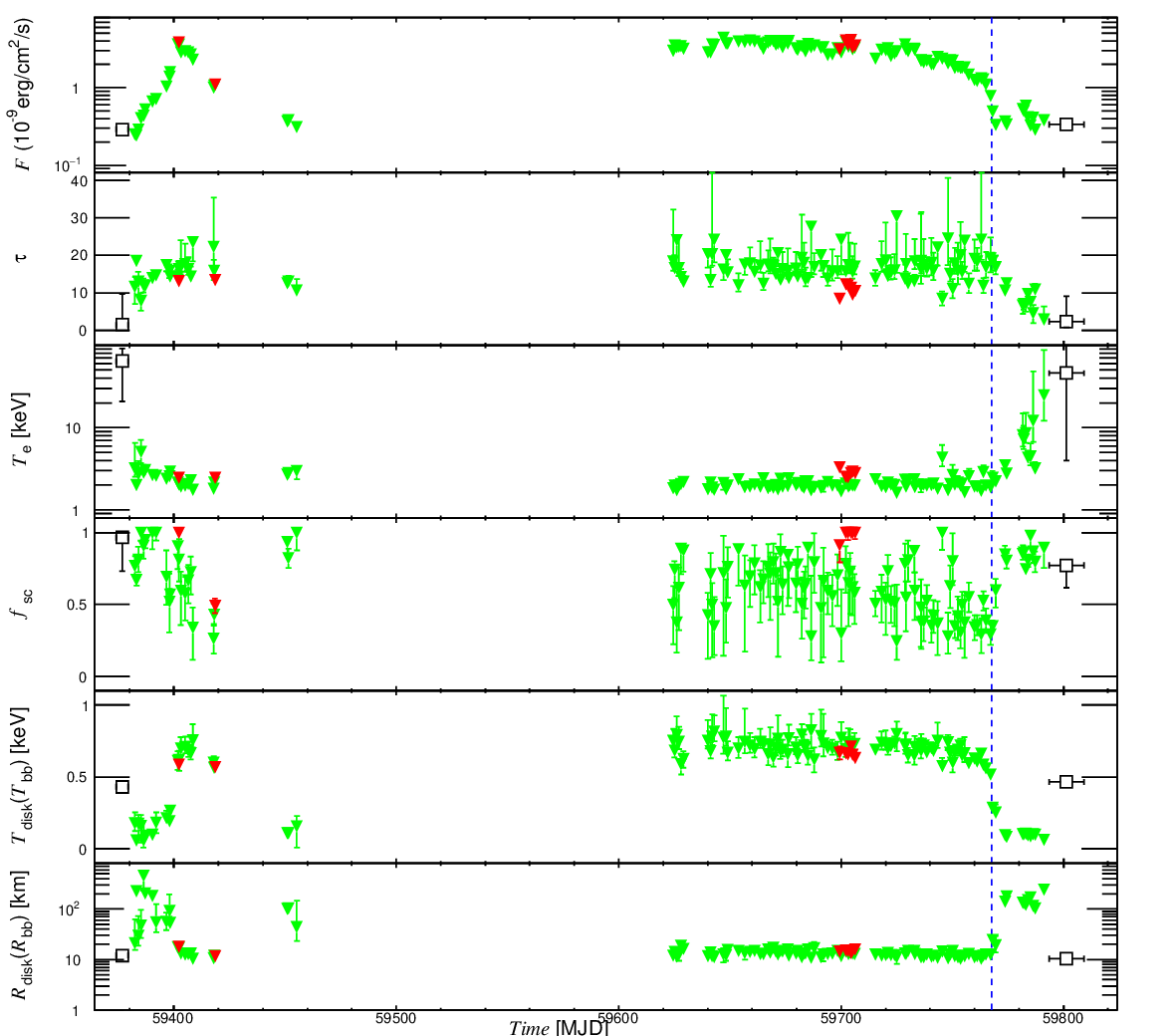}
 \caption{Time evolution of best-fitting parameters for the best-fitting model tbabs*thcomp*bbody (black) and tbabs*thcomp*(bbody+diskbb) (red and green) 
 from 4U~1730--22 during its  2021 and 2022 outbursts, including the bolometric flux $F$, the optical depth $\tau$, the electron temperature $T_{\rm e}$, the cover factor $f_{\rm sc}$, the accretion disk/blackbody temperature   $kT_{\rm disk}$/$kT_{\rm bb}$ and the inner disk radius/blackbody radius  $R_{\rm diskb}$/$R_{\rm bb}$ at   a distance of  7.56 kpc and in the face-on scenario (inclination angel $\theta$=0). 
The black markers represent the   parameters derived from  tbabs*thcomp*bbody in the onset and extinction time of the outbursts observed by NICER.  The green and  red  indicate the parameters derived from   tbabs*thcomp*(bbody+diskbb) (with {\bf fixed blackbody  parameters derived from the aforementioned spectral  fitting of the onset of the outburst)} observed by NICER and the jointed observations of NICER and Insight-HXMT, respectively. The 'propeller effect' time is marked by a blue line.  }
\label{fig_outburst_fit}
\end{figure}

\begin{figure}[t]
\centering
      \includegraphics[angle=0, scale=0.25]{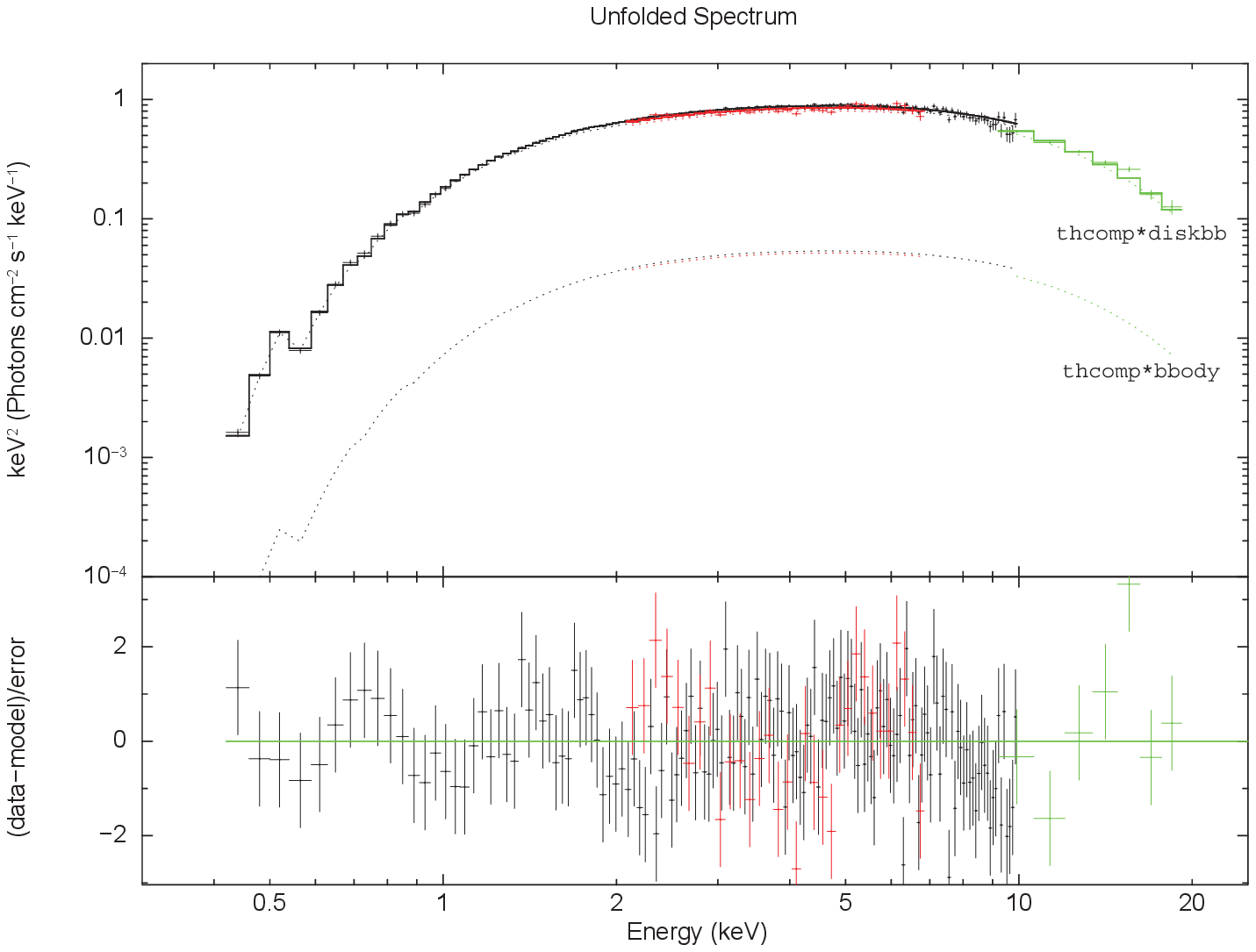}
            \includegraphics[angle=0, scale=0.25]{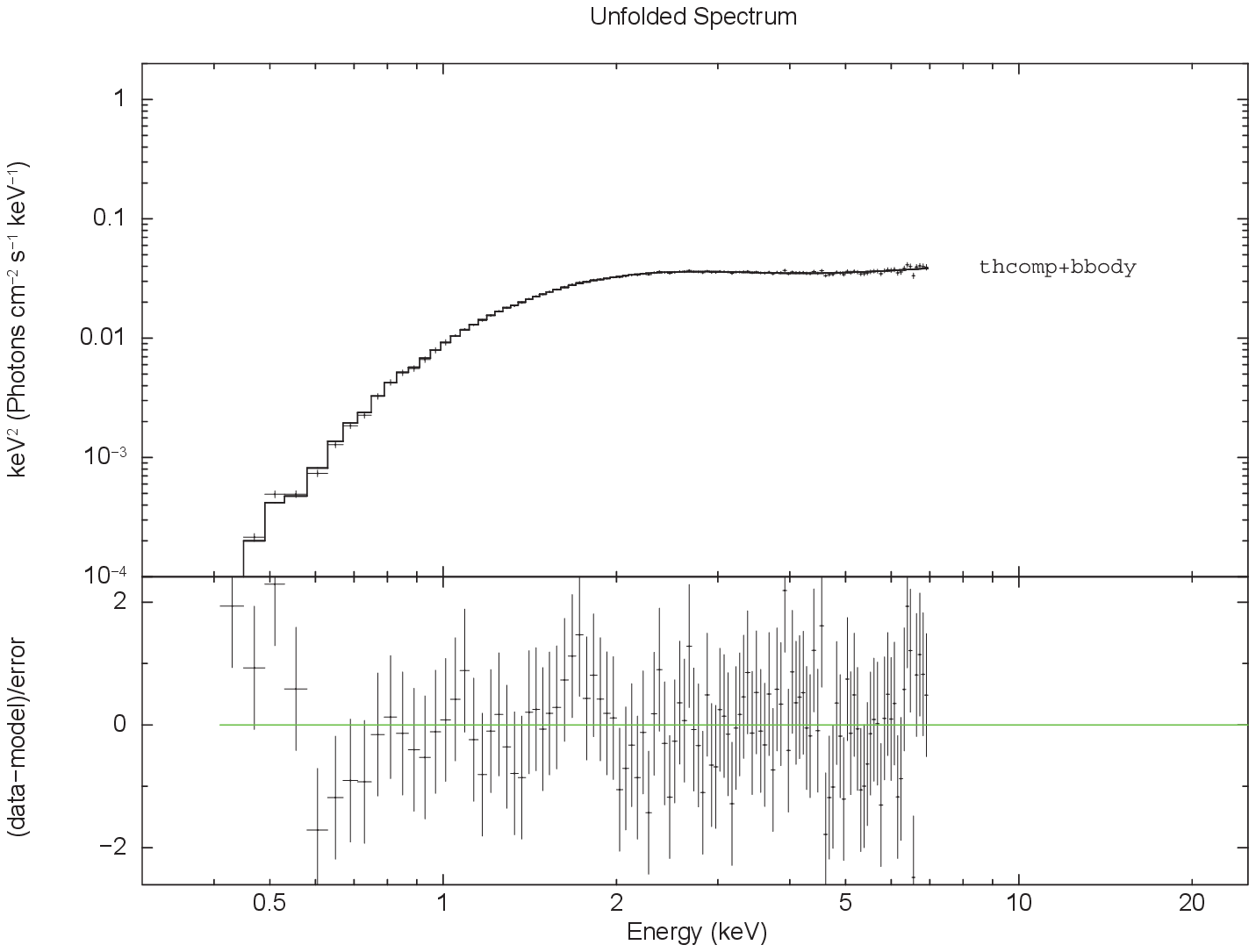}
  \caption{{\bf Left panel: the results of the spectral fits with data from NICER (black), LE (red), and ME (green) at the peak flux of the outburst in 2021 by cons*tbabs*thcomp*(bb+diskbb) with fixed blackbody parameters derived from the spectral fitting results of the onset of the outburst. Right panel:   the results of the spectral fits   with  NICER  data at the end of the outburst in 2022  by tbabs*thcomp*bb.}
 }
\label{fig_spec_residual_1}
\end{figure}

{\bf The obsids with thermonuclear bursts observed both by NICER and Insight-HXMT have been removed from the data before attempting spectral fits of the observations, since the additional thermal photons could  alter  the spectral fitting results of the persistent emission. These obsids all located at the peak phase of the outbursts, it does not affect the spectral evolution of the outbursts  when these obsids are removed. Moreover, some obsids with sharp variations in the lightcurves caused by the incorrect good-time-interval are also removed.} For the outburst onset of 2021 and extinction time of the outburst in 2022, i.e., the lowest count rates obsids ($\sim$ 30 cts/s) in Figure \ref{fig_hid}, the spectra of several obsids are stacked by the ftool addspec, respectively. We fit the two stacked spectra with an absorbed convolution thermal Comptonization model (with input photons contributed by the spectral component blackbody, {\bf i.e., tbabs*thcomp*bb}), available as thcomp (a more accurate version of nthcomp) \citep{Zdziarski2020} in XSPEC, which is described by the optical depth $\tau$, electron temperature $kT_{\rm e}$, scattered/covering fraction $f_{\rm sc}$.
The hydrogen column (tbabs in XSPEC) 
accounts for both the line-of-sight column density and any intrinsic absorption near the source.
The seed photons are in the shape of blackbody since the thcomp model is a convolution model, and a fraction of Comptonization photons are also given in the model.
The results of the spectral fitting are presented in Table \ref{tb_fit_thcomp_bb} and plotted in Figure \ref{fig_outburst_fit} and Figure \ref{fig_spec_residual_1}. The inferred blackbody radius is well consistent with the NS radius under a distance of 7.56 kpc \citep{Li2022}, thus a spherical corona scenario is favored since the scattered/covering fraction $f_{\rm sc}>$ 70\% at this stage.
We also notice that the blackbody temperature of the two stacked spectra are both around 0.45 keV and the bolomeric fluxes of them are also very similar.

Along with the increase of the count rates, the model above is also attempted to fit the obsids with count rate $>$ 40. However, the parameters derived from the above model is unreasonable, e.g., the blackbody radius $R_{\rm bb}$ is much larger than 10 km and up to more than 100 km, which is much larger than the NS radius. Moreover, the blackbody temperature $kT_{\rm bb}$ decreases as the count rate increases, which is also very unlikely.
For the obsids in the rising phase of the outburst in 2021 with luminosity (4.5--13.8$\times10^{-9}~{\rm erg/cm}^{2}/{\rm s}$) and count rate (70--190 cts/s) which are similar to that of the two obsid of the soft-to-hard state transition (see the next paragraph), the inferred $kT_{\rm bb}$ is 0.39--0.11 keV and 20--120 km.
Under this condition, the spherical corona scenario is disfavored during the rising phase of the outburst.
Thus, we assume the NS surface emission (the temperature and the area) does not change during the outbursts, and take it as part of the seed photons of the Comptonization, {\bf i.e., tbabs*thcomp*(bb+diskbb) with the fixed blackbody parameters derived from the spectral fitting at the outburst onset.}
This assumption should underestimate the NS surface emission and thus leads to an overestimate the accretion disk emission/radius, since $kT_{\rm bb}$ should be higher as the accretion rate increases.
The underestimation of the inclination angle (we take $\theta$ = 0 to calculate the inner disk radius) could offset the overestimation above of the disk emission/radius. 
Some works \citep{Thompson2005,Zand2009} indicated that the NS surface temperature could be up to 0.6--0.8 keV and partial of the NS surface emission is blocked by the disk \citep{Thompson2005} and not involved in the Comptonization, which could also offset the overestimate the accretion disk emission/radius. 
As given in the following paragraph, 
the derived inner radii are consistent well with the NS radius, which indicates this assumption has a mild influence on the spectral fitting and is accepted. 

{\bf For consistency check, we also fit the spectra by removing the NS surface emission, the derived model parameters are shown in the Appendix 
both for the tables and figure.  
For obsids around the peak of the outbursts (e.g., with count rates $>$ 200 cts/s), it has an insignificant influence on the derived parameters of the thcomp and diskbb, since the flux of the disk is more then ten times higher than the blackbody emission. However, for the obsids around the rising and decaying parts of the outburst (e.g., with count rates $<$ 100 cts/s), the influence is significant since the flux of the disk is comparable with the blackbody emission; e.g. for the obsids 4639010182--4639010187 around the propeller effect, the derived $T_{\rm disk}$ changes from 0.6 to 1.0 keV and  $R_{\rm disk}$ changes from 10 to 2 km, which is unlikely because of the non-physical temperature trend and the small inner disk radius. Moreover, the model of the spherical corona scenario (tbabs*thcomp*bb) is also attempted to fit obsids 4639010182--4639010187 around this transition, and the derived $T_{\rm bb}$ changes from 0.35 keV to 0.48 keV and $R_{\rm bb}$ changes from 45 km to 16 km, which is also unlikely because of the non-physical trend of the temperature and the large NS radius. After the transition, e.g., the obsids 4639010188--4639010196  with count rates $\sim$ 52--40 cts/s, the derived $T_{\rm bb}$ changes from 0.40 keV to 0.35 keV and $R_{\rm bb}$ changes from 16 km to 11 km in the spherical corona scenario, which indicates that the data quality of these obsids could not identify whether the spherical corona or the disk corona scenario is more reasonable. Given the overall consistency and reasonable evolution of model parameters of the model including both components of the NS surface and accretion disk emission, we take this model as physically more appropriate than that with only one thermal emission component.
}

\begin{figure}[t]
\centering
\includegraphics[angle=0, scale=0.25]{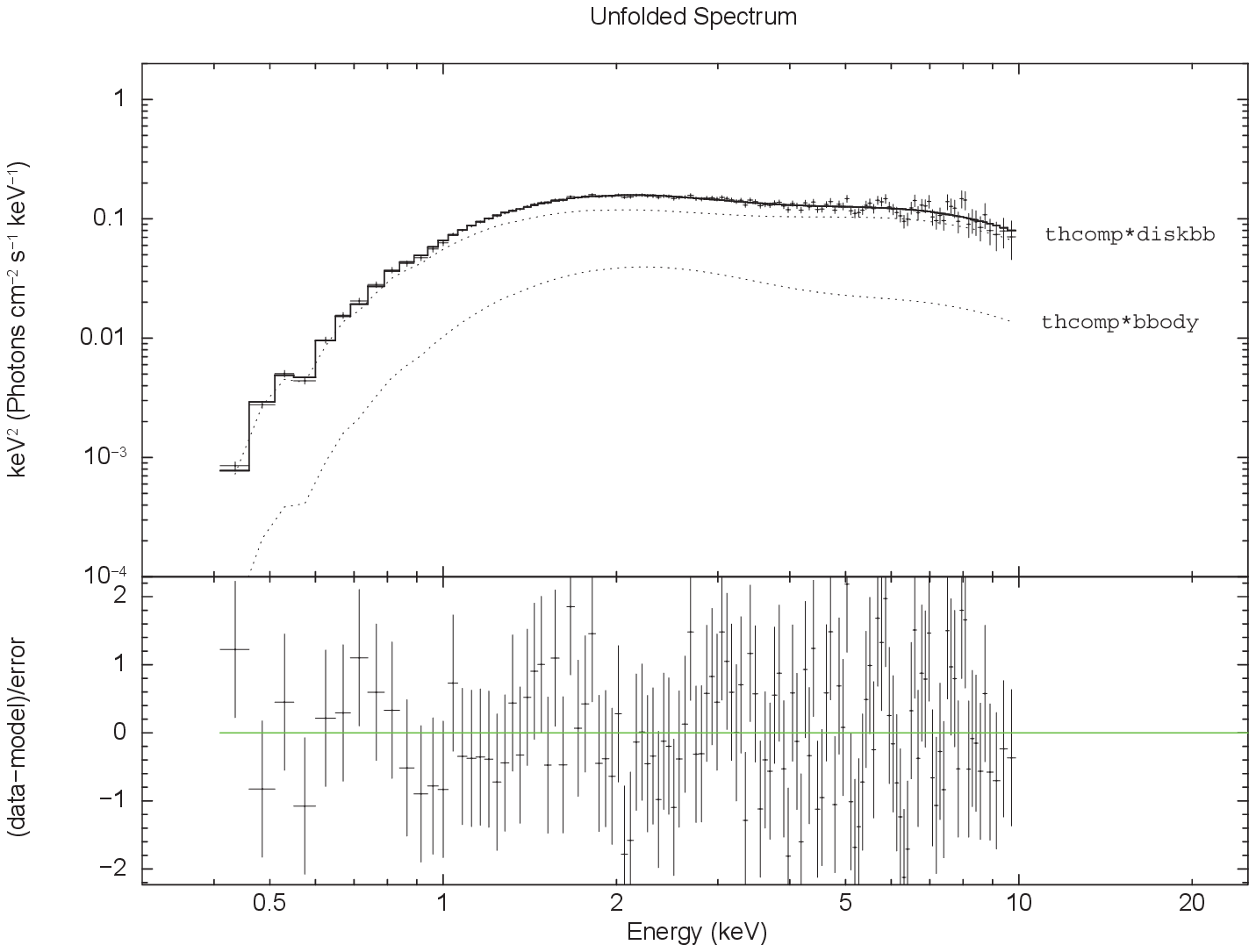}
\includegraphics[angle=0, scale=0.25]{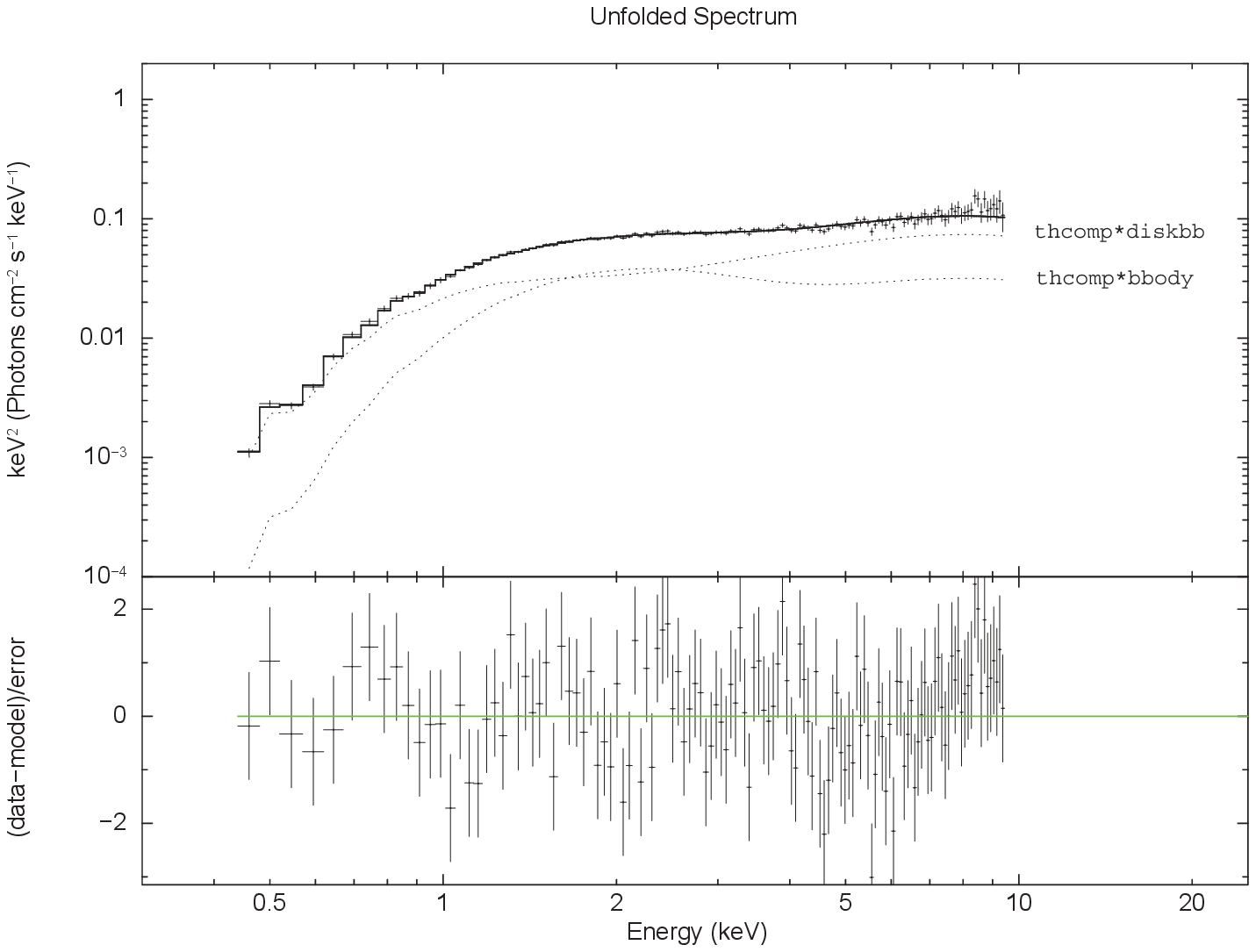}
 \caption{{\bf The results of the spectral fits of the obsid 4639010184 (left) and  obsid 4639010185 (right),} i.e., the  two obsids at the transition time when the 'propeller effect' occurred at the decay phase of the outburst in 2022  by tbabs*thcomp*(bb+diskbb).
 }
\label{fig_spec_residual_2}
\end{figure}

We then take the thermal emission from the accretion disk as the other part of the seed photons {\bf to fit the obsids with count rates $>$ 40 cts/s}, i.e., the model is revised to tbabs*thcomp*(bb+diskbb) {\bf with the fixed blackbody parameters derived from the spectral fitting at the outburst onset}.
From the revised model, the derived parameters are reasonable, as shown in Table \ref{tb_fit_thcomp_bb_diskbb}, Figure \ref{fig_outburst_fit} and Figure \ref{fig_spec_residual_2}.
{\bf Please note that the error bars of the disk temperature $T_{\rm disk}$, inner disk radii $R_{\rm disk}$ and the bolometric fluxes $F$ of some  obsids (mostly with counts rates $<$ 60 ct/s) with $T_{\rm disk}\sim$ 0.1--0.2 keV is hard to be derived because of the low temperature and low flux of the disk emission, and we fixed other parameters to calculate $T_{\rm disk}$, $R_{\rm disk}$, and $F$, which should underestimate these error bars. We also attempt to stack several of them, and the derived parameters are consistent with that derived from the individual obsid. }
The model is also used to fit the joint spectra of NICER and Insight-HXMT,  {\bf as shown in Table \ref{tb_fit_thcomp_bb_diskbb_nicer_hxmt} and Figure \ref{fig_outburst_fit}.}
Under a distance of 7.56 kpc, and $L_{\rm Edd}=1.8\times10^{38}$ erg/s, the peak luminosity 
corresponds to $\sim$15\% $L_{\rm Edd}$.
Along with the increased flux, at the rising phase of the outburst, the inner disk radius $R_{\rm disk}$ is above 20 km, and then stays at $\sim$ 10 km after the flux peak time in the face-on scenario (inclination angel $\theta$ = 0). 
At the decay phase of the outburst in 2022, an abrupt change of $R_{\rm disk}$ and $T_{\rm disk}$ between obsids 4639010184 and 4639010185 is obvious, i.e., from 0.5 keV and 12 km to 0.3 keV and 24 km within one day.
NICER missed the decay phase of the outburst in 2021 with a gap about 40 days.
Please note that there are some differences of the thcomp parameters derived from the joint NICER/Insight-HXMT spectral fitting and the NICER spectral fitting, but the trends are very similar.
It is because that the shape of the unsaturated Comptonization is more dependant on the higher energy photons which undergo more scatterings than the mean or the lower energy photons. The joint spectral fitting extends the photon energy from 10 keV to 20 keV, and the derived thcomp parameters should be more reliable.
We also notice that the disk parameters derived from the two kinds of spectra are consistent well with each other.

Moreover, the enlarged $R_{\rm disk}$ is consistent well with the corotation radius $r_{\rm c}$, both of which are around 24 km.
The transition is also obvious in the HID, as shown in Figure \ref{fig_hid}, with count rates from 174 cts/s to 85 cts/s within one day.
We also notice that, different from the bolometric luminosity of the disk $F_{\rm disk}$ which decreases by a factor of 2.9 during the transition, however, the parameters of the Comptonization model changed little. This leads to that the total bolometric luminosity shows a mild change by a factor of 1.6. 


Under a distance of 7.56 kpc, the transition occurred is between obsids 4639010184 and 4639010185 in Table \ref{tb_fit_thcomp_bb_diskbb}, with the bolometric luminosities of $5.4\times10^{36}~{\rm erg}~{\rm s}^{-1}$ and $3.4\times10^{36}~{\rm erg}~{\rm s}^{-1}$, respectively.
Assuming $M_{\rm NS}=$1.4 $M_{\odot}$, $R_{\rm NS}$=10 km, and taking $P_{\rm spin}$=1/587=1.71 ms, the inferred magnetic fields corresponding to the two luminosities above are $1.8\times10^{8}$ G and $2.2\times10^{8}$ G, respectively, based on the formula given by \citet{Lamb1973, Cui1997, Zhang1998} 
\begin{eqnarray}
 L_{X,36}\approx2.34B_{9}^{2}P_{-2}^{-7/3}M_{1.4}^{-2/3}R_{6}^{5},
   \label{magn}
\end{eqnarray}
here $L_{X,36}$ is the total luminosity in units of $10^{36}~{\rm erg}~{\rm s}^{-1}$, $B_{9}$ is the magnetic field in units of $10^{9}$ G, $P_{-2}$ is the NS period in units of 10 ms, $M_{1.4}$ is the NS mass in units of 1.4$M_{\odot}$ and $R_{6}$ is the NS radius in units of 10 km.

\section{Discussion}



Different from the previous work on the propeller effect, we take both the NS surface and accretion disk emission as the seed photons of the Comptonization and get the time evolution of the corona and disk in the bolometric luminosity range of 1\%--15\% $L_{\rm Edd}$. The HID shows a hysteresis, i.e., the luminosity of the hard-to-soft transition is higher than the luminosity of the soft-to-hard transition. 
However, the first transition above (hard-to-soft) was not observed by NICER due to its observation gap.
Fortunately, the last transition above (soft-to-hard) was detected by NICER and shows a sudden decline in accretion disk emission.
The truncated inner disk radius is consistent well with the corotation radius.
Taking the soft-to-hard transition as caused by the propeller effect, the magnetic field is derived.
Apart from Aql~X--1 \citep{Zhang1998} and 4U~1608--52 \citep{Chen2006}, 4U~1730--22 is now also listed among the sources with observed propeller effect.

\begin{figure}[t]
\centering
      \includegraphics[angle=0, scale=0.10]{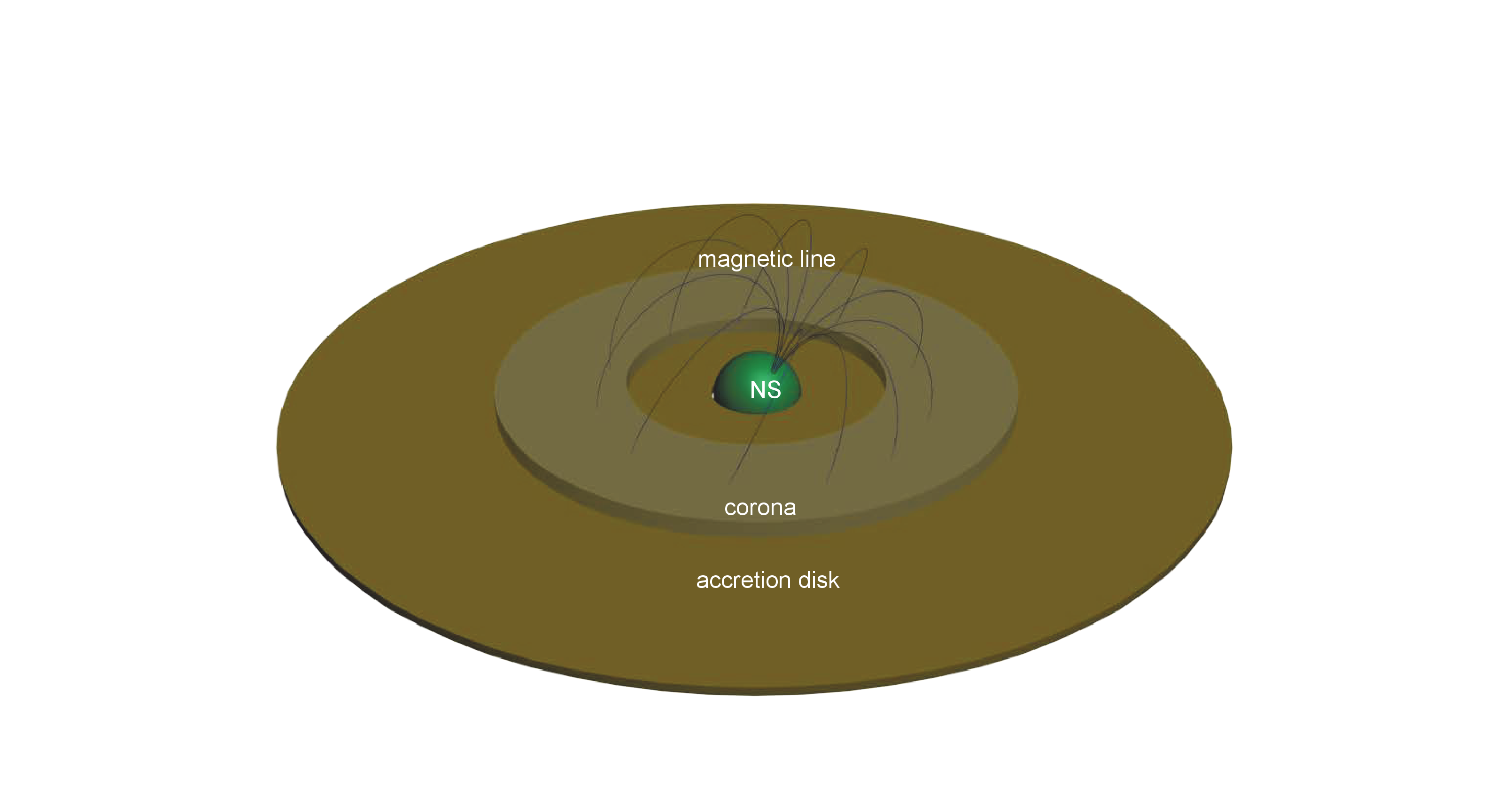}
            \includegraphics[angle=0, scale=0.10]{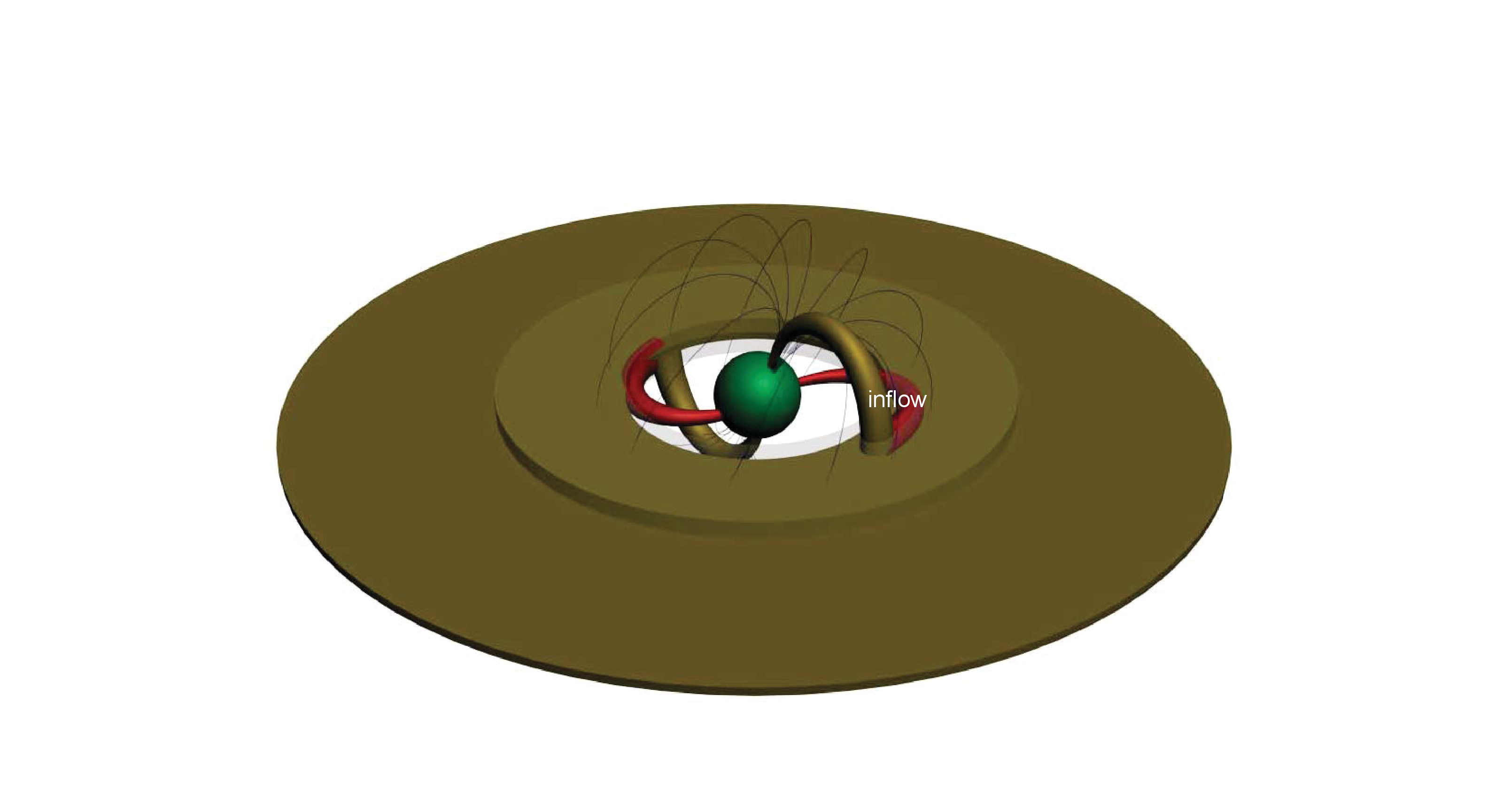}
      \includegraphics[angle=0, scale=0.10]{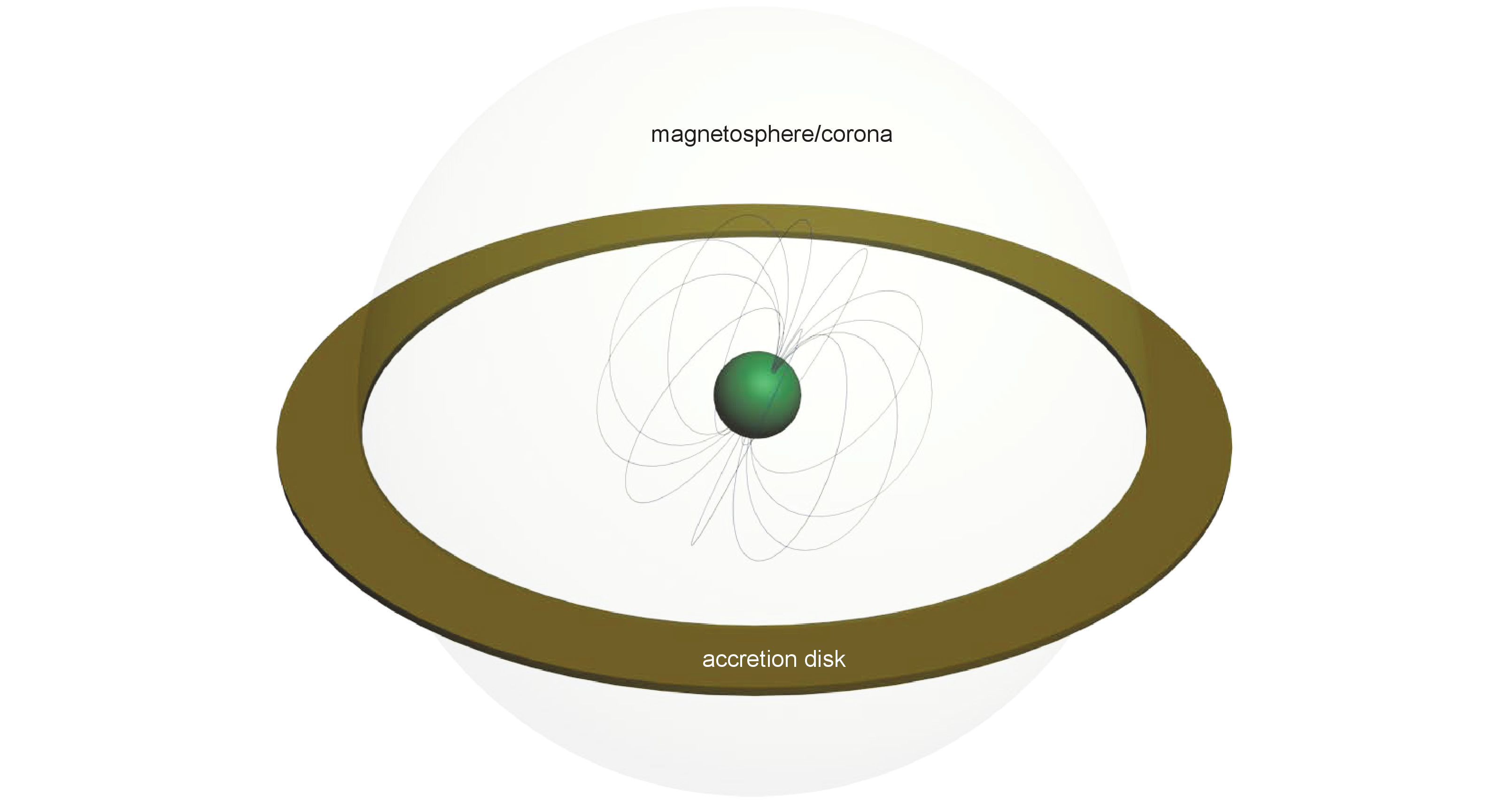}
 \caption{Illustration of the central region of an NS LMXB in the high/soft state (left, before the propeller effect), just after the propeller effect (middle) and the tail of the outburst (right).
 Before the propeller effect, the accretion disk extends to the NS surface; and just after that the accretion disk is truncated near the corotation radius, and partial material leaks to the NS surface; at the end of the outburst, there is a scenario of the  magnetospheric accretion and the spherical corona.
 }
\label{fig_illustraction}
\end{figure}

We notice that the total luminosity change during the transition is by a factor of 1.6, which is smaller than the model-predicted luminosity gap of 2.4, and could be related to matter leaking through the magnetosphere to the NS surface (e.g., \citealp{Stella1986,Zhang1998}), e.g., forming some kind of accretion flow/channel to the NS surface (e.g., magnetic poles) \citep{Arons1976, Elsner1977}, as shown in Figure \ref{fig_illustraction}. If it were true, a reasonable prediction is that there is an enhancement of pulsation fraction (if the pulsation exists) around the transition, which is beyond scope of this work and will be explored elsewhere.



For the outbursts of 4U~1730--22, the picture aforementioned in Section 1 is revised as shown in the toy model of Figure \ref{fig_illustraction}.
In the high/soft state, the accretion disk extends to the NS surface. Along with the accretion rate to a certain value, 
the propeller effect happens and expels out most material within the magnetosphere, which leads to a truncated disk (at the corotation radius for this work). 
However, there is still partial material that could be leaked to the NS surface. At end of the outburst, the inner radius of the accretion disk and the radius of the magnetosphere is far from the NS, i.e., the spherical corona scenario.


Particularly, we notice that the two luminosities above are roughly the same as the transition luminosities detected by RXTE in 2004 from 4U~1608--52 \citep{Chen2006}, e.g., $5.3\times10^{36}~{\rm erg}~{\rm s}^{-1}$ and $3.3\times10^{36}~{\rm erg}~{\rm s}^{-1}$.
Considering that the spin periods of the two systems are similar, e.g., the spin period of 4U~1608--52 is 1.16 ms, the behavior of the outburst should resemble each other.
The similar outburst's behavior is reminiscent of the similar thermonuclear bursts' (type I X-ray bursts) behavior of the two systems \citep{Chen2022a,Chen2022b}, e.g., the bright photospheric radius expansion (PRE) bursts show a shortage during the rising PRE phase which could be due to the occlusion by the disk which is close to the NS surface during their high/soft state.

{\bf Moreover, the luminosity gap ratio of `2.4' works only at the aligned case, i.e., aligned magnetic and spin axes, which are both perpendicular to the plane of the disk; a large inclination angle between the NS rotation and magnetic field axes leads to a smaller gap ratio, from a mechanism that has been supported
by theory and simulations several works (e.g., \citealp{2009A&A...493..809B,2015SSRv..191..339R,2018A&A...617A.126B}). In this quasi-magnetospheric accretion scenario, if the matter accumulates around the magnetospheric radius faster than it can be ejected, the episodic or cyclic outbursts could occur \citep{2012MNRAS.420..416D,2014MNRAS.441...86L}. We notice that after this work, 4U~1730--22 started a new outburst in September and October 2022, which implies this source stepped into an active phase and could be related the mass accumulation mentioned above.  }

Both for black hole (BH)-LMXBs and NS-LMXBs, around the soft-to-hard state transition, the luminosity shows a fast decay stage, which is the `knee' feature \citep{Powell2007}.
For the NS-LMXBs, it could be related to the propeller effect. 
However, for BH-LMXBs, the propeller effect is not expected in absence of magnetic field, another scenario--the thermal disk instability could be related to the `knee' feature, which could be also related to the hard-to-soft transition
\citep{maccarone2003}. In the rising phase of the outburst in 2021 from 4U~1730--22, we notice that at the same luminosity of the soft-to-hard state transition, the transition of the hard-to-soft state did not happen, but could have occurred at a higher luminosity and not observed due to the observation gap, i.e., the hysteresis \citep{Fender2004}. 
\citet{maccarone2003} proposed that the transitions between the optically thick and optically thin corona are related to the hard-to-soft and soft-to-hard transitions, and thus the propeller effect is not the sole cause of the state transitions.

The above calculation and estimation are based on the static analysis and do not consider the radial velocity component of the innermost accretion disk, i.e., the inertia of the disk.
In the rising phase of the outburst, the accretion disk has a radial velocity towards the NS and the inflow is easier to overcome the magnetic stress; in contrast, in the decay phase the accretion disk has a radial velocity outwards from the NS and the inflow is harder to overcome the magnetic stress. In other words, the higher inflow rate of the rising phase could fill the luminosity gap caused by the propeller factor more significantly than that of the decay phase.
The thought above is only a qualitative analysis, and a more precise quantitative analysis should be explored further.

Nevertheless, the corona parameters ($\tau$, $kT_{\rm e}$ and $f_{\rm sc}$) change little during the soft-to-hard transition, although the inner part of the corona (from the NS surface to the corotation radius) has been expelled out the system.
This finding implies that the structure and property of the corona have little dependence on the radial distance, and 
the whole distributing of the corona is stable in this region.
Since the disk viscous power is a strong function of the radial distance and decays rapidly outwards, this stable corona is against the origin that the corona's power comes from the thermal electrons from the disk's viscous power.
Thus, another origin of the corona--the magnetic field is favored, and it could be the reservoir depositing much more energy than the thermal energy content \citep{Merloni2001}.



In theory, the NS surface temperature is expected to increase as the accretion rate increases.
However, the above results are obtained by the assumption that the NS surface emission has a mild change during the whole outburst.
The small counts of the spectrum prevent us unfixing the NS surface parameters, due to the effective area is not big enough to distinguish the accretion disk and the NS surface emission.
A larger detection area and broadband energy coverage may be satisfied by the next generation of Chinese mission of so-called eXTP (enhanced X-ray Timing and Polarimetry mission) \citep{Zhang2019}. 

\acknowledgements
This work made use of the data and software from the Insight-HXMT
mission, a project funded by China National Space Administration
(CNSA) and the Chinese Academy of Sciences (CAS).
This research has  made use of data and software provided by of data obtained from the High Energy Astrophysics Science
Archive Research Center (HEASARC), provided by NASA’s
Goddard Space Flight Center.
This work is supported by the National Key R\&D Program of China (2021YFA0718500) and the National Natural Science Foundation of China under grants 11733009, U1838201, U1838202, U1938101, U2038101, 12130342, U1938107, 12173103.
This work was partially supported by International Partnership Program of Chinese Academy of Sciences (Grant No.113111KYSB20190020).

\bibliographystyle{plainnat}


\clearpage

\begin{table}\tiny
\centering
\caption{The results of the spectral fit of the  NICER spectra in the 0.4--7 keV range   with  cons*tbabs*thcomp*bb at the onset of the outburst in 2021 and the extinction time of the outburst in 2022}
\label{tb_fit_thcomp_bb}
\begin{tabular}{ccccccccccccc}
\\\hline 
No &Time &$N_{\rm H}$  & $\tau$ & $kT_{\rm e}$  & $f_{\rm sc}$ & $kT_{\rm bb}$ & $R_{\rm bb}$ & $F_{\rm bb}$ &$F_{\rm corona}$ &$F_{\rm total}$ & $\chi_{\nu}^{2}$ (d.o.f.)\\
   & MJD & $10^{22}~{\rm cm}^{-2}$ &  & keV & &keV & km & $10^{-9}$  & $10^{-9}$  & $10^{-9}$ & \\
\hline
1   & 59376.74    & $0.39\pm0.009$  & $1.5_{-0.2}^{+8.2}$ & $64.01_{-43.19}^{+27.39}$ & $0.97_{-0.2}^{}$ & $0.43_{-0.01}^{+0.01}$ & $12.0_{-0.4}^{+0.4}$ & $0.094_{-0.001}^{+0.001}$ & $0.195_{-0.011}^{+0.008}$ & $0.290_{-0.011}^{+0.008}$ & $0.65(136)$ \\\hline
2  & 59801.32    & $0.39\pm0.005$  & $2.3_{-1.6}^{+6.7}$ & $45.82_{-41.86}^{+179.90}$ & $0.77_{-0.2}^{}$ & $0.47_{-0.01}^{+0.01}$ & $10.4_{-0.2}^{+0.2}$ & $0.095_{-0.001}^{+0.001}$ & $0.240_{-0.001}^{+0.001}$ & $0.335_{-0.001}^{+0.001}$ & $0.74(109)$ \\\hline
\end{tabular}
\begin{list}{}{}
\item[]{{\bf The onset of the outburst 2011  includes the NICER obsids 4202200101--42002200106, the extinction time of the outburst in 2022 includes the NICER obsids 4639010197--4639010210, as shown in the green boxes in Figure \ref{fig_lc_nicer_le_me} and Figure \ref{fig_hid}. } The model parameters: the optical depth $\tau$,  the electron temperature $kT_{\rm e}$,  the cover factor $f_{\rm sc}$,  the blackbody  temperature   $kT_{\rm bb}$ and  the blackbody radius   $R_{\rm bb}$ at a distance of 7.56 kpc, the bolometric flux of the blackbody $F_{\rm bb}$, the bolometric flux of the corona $F_{\rm bb}$, and the total  bolometric flux  $F_{\rm total}$ are in units of $10^{-9}~{\rm erg/cm}^{2}/{\rm s}$, reduced $\chi_{\nu}^{2}$ and the degree of freedom (d.o.f.). Please note the energy range of the two stacked spectra--the onset of the outburst in 2011 and the extinction time of the outburst in 2022 (right) are truncated to 0.4--7 keV because of the high background level at $>$ 7 keV. }
\end{list}
\end{table}

\clearpage
\begin{landscape}

\begin{table}\tiny
\centering
\caption{The results of the spectral fit of the jointed observations of  NICER, Insight-HXMT/LE, and Insight-HXMT/ME  in the 0.4--20 keV range   with  cons*tbabs*thcomp*(bb+diskbb) during the outbursts in 2021 and 2022}
\label{tb_fit_thcomp_bb_diskbb_nicer_hxmt}
\begin{tabular}{cccccccccccccc}
\\\hline 
No & HXMT obsid & Time &NICER obsid &$N_{\rm H}$  & $\tau$ & $kT_{\rm e}$  & $f_{\rm sc}$ & $kT_{\rm disk}$ & $R_{\rm disk}$ & $F_{\rm disk}$ &$F_{\rm corona}$ &$F_{\rm total}$ & $\chi_{\nu}^{2}$ (d.o.f.)\\
   & & MJD & & $10^{22}~{\rm cm}^{-2}$ &  & keV & &keV & km & $10^{-9}$  & $10^{-9}$  & $10^{-9}$ & \\\hline
 1 &P041401100101-20210707-01-01 & 59402.25 & 4202200123   & $0.52\pm0.006$  & $13.1_{-0.2}^{+0.3}$ & $2.47_{-0.06}^{+0.05}$ & $1.00_{-0.02}^{}$ & $0.59_{-0.02}^{+0.02}$ & $17.9_{-0.9}^{+1.1}$ & $1.435_{-0.007}^{+0.009}$ & $2.298_{-0.021}^{+0.020}$ & $3.823_{-0.020}^{+0.018}$ & $1.05(181)$ \\\hline
2 &P041401101101-20210723-01-01 & 59418.48 & 4202200131   & $0.52\pm0.008$  & $13.5_{-1.1}^{+1.3}$ & $2.45_{-0.16}^{+0.19}$ & $0.49_{-0.05}^{+0.05}$ & $0.57_{-0.02}^{+0.02}$ & $11.6_{-0.8}^{+0.9}$ & $0.528_{-0.005}^{+0.005}$ & $0.477_{-0.011}^{+0.011}$ & $1.095_{-0.010}^{+0.010}$ & $0.90(170)$ \\\hline
3 &P051400200103-20220430-01-01 & 59699.41 & 4639010134   & $0.53\pm0.009$  & $8.4_{-1.0}^{+1.0}$ & $3.31_{-0.36}^{+0.49}$ & $0.91_{-0.12}^{}$ & $0.67_{-0.05}^{+0.04}$ & $14.5_{-1.4}^{+1.7}$ & $1.591_{-0.014}^{+0.014}$ & $1.433_{-0.036}^{+0.041}$ & $3.114_{-0.033}^{+0.039}$ & $0.80(161)$ \\\hline
4 &P051400200401-20220503-01-01 & 59702.08 & 4639010135   & $0.52\pm0.004$  & $12.2_{-0.2}^{+0.4}$ & $2.55_{-0.07}^{+0.07}$ & $1.00_{-0.02}^{}$ & $0.66_{-0.02}^{+0.02}$ & $15.2_{-0.7}^{+0.8}$ & $1.684_{-0.008}^{+0.009}$ & $2.306_{-0.019}^{+0.028}$ & $4.080_{-0.017}^{+0.026}$ & $1.02(188)$ \\\hline
5 &P051400200501-20220504-01-01 & 59703.07 & 4639010136   & $0.52\pm0.005$  & $12.0_{-0.5}^{+0.5}$ & $2.49_{-0.09}^{+0.10}$ & $1.00_{-0.05}^{}$ & $0.66_{-0.02}^{+0.03}$ & $14.8_{-1.0}^{+0.8}$ & $1.538_{-0.008}^{+0.009}$ & $1.944_{-0.022}^{+0.019}$ & $3.572_{-0.020}^{+0.017}$ & $1.27(184)$ \\\hline
6 &P051400200602-20220505-01-01 & 59704.44 & 4639010137   & $0.51\pm0.004$  & $11.3_{-0.3}^{+0.3}$ & $2.80_{-0.07}^{+0.07}$ & $1.00_{-0.01}^{}$ & $0.71_{-0.02}^{+0.02}$ & $13.5_{-0.5}^{+0.5}$ & $1.751_{-0.007}^{+0.007}$ & $2.299_{-0.020}^{+0.024}$ & $4.140_{-0.019}^{+0.023}$ & $1.14(188)$ \\\hline
7 &P051400200702-20220506-01-01 & 59705.20 & 4639010138   & $0.52\pm0.004$  & $9.6_{-0.4}^{+0.3}$ & $2.93_{-0.09}^{+0.13}$ & $1.00_{-0.01}^{}$ & $0.65_{-0.01}^{+0.01}$ & $15.2_{-0.5}^{+0.5}$ & $1.579_{-0.007}^{+0.007}$ & $1.679_{-0.017}^{+0.017}$ & $3.348_{-0.015}^{+0.016}$ & $1.11(192)$ \\\hline
8 &P051400200802-20220507-01-01 & 59706.29 & 4639010139   & $0.53\pm0.004$  & $10.6_{-0.3}^{+0.4}$ & $2.81_{-0.10}^{+0.10}$ & $1.00_{-0.04}^{}$ & $0.63_{-0.01}^{+0.02}$ & $16.0_{-0.9}^{+0.7}$ & $1.526_{-0.008}^{+0.009}$ & $1.872_{-0.024}^{+0.019}$ & $3.488_{-0.022}^{+0.017}$ & $1.18(187)$ \\\hline
   \end{tabular}
\begin{list}{}{}
\item[a]{: The model parameters: the optical depth $\tau$,  the  electron temperature $kT_{\rm e}$,  the cover factor $f_{\rm sc}$,  the accretion disk  temperature   $kT_{\rm disk}$ and  the inner disk radius   $R_{\rm diskb}$ at a distance of 7.56 kpc and in the face-on scenario (inclination angel $\theta$=0), the bolometric flux of the diskbb $F_{\rm diskbb}$, the bolometric flux of the corona $F_{\rm diskbb}$, and the total  bolometric flux  $F_{\rm total}$ are in units of $10^{-9}~{\rm erg/cm}^{2}/{\rm s}$, reduced $\chi_{\nu}^{2}$ and the degree of freedom (d.o.f.).}
\end{list}
\end{table}

\clearpage

\begingroup\tiny
\begin{longtable}{ccccccccccccccc}
\caption{The results of the spectral fit of the  NICER spectra in the 0.4--10 keV range with  tbabs*thcomp*(bb+diskbb)}
\label{tb_fit_thcomp_bb_diskbb}
\\\hline 
No & obsid & Time & Exposure & Rate &$N_{\rm H}$  & $\tau$ & $kT_{\rm e}$  & $f_{\rm sc}$ & $kT_{\rm disk}$ & $R_{\rm disk}$ & $F_{\rm disk}$ &$F_{\rm corona}$ &$F_{\rm total}$ & $\chi_{\nu}^{2}$ (d.o.f.)\\
   & &MJD & s & cts/s & $10^{22}~{\rm cm}^{-2}$ &  & keV & &keV & km & $10^{-9}$  & $10^{-9}$  & $10^{-9}$ & \\\hline
   \endhead
 1&4202200109&59382.45&2700&43.5 &$0.46\pm0.021$&$11.7_{-4.6}^{+2.6}$&$3.18_{-0.64}^{+3.32}$&$0.77_{-0.06}^{+0.07}$&$0.18_{-0.06}^{+0.07}$&$21.0_{-5.6}^{+41.1}$&$0.015_{-0.001}^{+0.001}$&$0.144_{-0.001}^{+0.001}$&$0.249_{-0.001}^{+0.001}$&$0.64(121)$ \\\hline
2&4202200110&59383.23&2430&46.8 &$0.43\pm0.005$&$18.5_{-1.1}^{+1.3}$&$2.02_{-0.08}^{+0.08}$&$0.67_{-0.04}^{+0.04}$&$0.06_{-0.01}^{+0.01}$&$223.1_{-2.3}^{+2.3}$&$0.012_{-0.001}^{+0.001}$&$0.140_{-0.001}^{+0.001}$&$0.243_{-0.001}^{+0.001}$&$0.84(123)$ \\\hline
3&4202200111&59384.00&2062&48.9 &$0.50\pm0.020$&$12.9_{-3.3}^{+2.7}$&$2.72_{-0.44}^{+1.19}$&$0.81_{-0.08}^{+0.09}$&$0.17_{-0.12}^{+0.06}$&$29.7_{-10.9}^{+43.2}$&$0.027_{-0.001}^{+0.001}$&$0.169_{-0.001}^{+0.001}$&$0.285_{-0.001}^{+0.001}$&$1.18(118)$ \\\hline
4&4202200112&59385.16&5585&58.1 &$0.54\pm0.035$&$7.9_{-2.6}^{+2.1}$&$5.06_{-1.38}^{+2.04}$&$1.00_{-0.06}^{}$&$0.16_{-0.12}^{+0.08}$&$47.3_{-20.4}^{+49.2}$&$0.047_{-0.001}^{+0.001}$&$0.267_{-0.001}^{+0.001}$&$0.404_{-0.001}^{+0.001}$&$0.83(133)$ \\\hline
5&4202200113&59386.33&2544&67.4 &$0.54\pm0.020$&$12.1_{-2.0}^{+1.8}$&$2.86_{-0.38}^{+0.67}$&$0.91_{-0.09}^{+0.04}$&$0.06_{-0.05}^{+0.13}$&$457.3_{-225.2}^{}$&$0.057_{-0.001}^{+0.001}$&$0.291_{-0.001}^{+0.001}$&$0.438_{-0.001}^{+0.001}$&$1.07(125)$ \\\hline
6&4202200114&59387.11&5027&76.7 &$0.57\pm0.009$&$11.9_{-1.0}^{+1.0}$&$3.05_{-0.26}^{+0.36}$&$0.95_{-0.06}^{}$&$0.09_{-0.01}^{+0.01}$&$206.2_{-0.5}^{+0.5}$&$0.073_{-0.001}^{+0.001}$&$0.364_{-0.001}^{+0.001}$&$0.528_{-0.001}^{+0.001}$&$0.94(134)$ \\\hline
7&4202200115&59390.26&1118&96.3 &$0.56\pm0.010$&$13.9_{-0.4}^{+1.4}$&$2.65_{-0.25}^{+0.13}$&$1.00_{-0.12}^{}$&$0.10_{-0.01}^{+0.01}$&$180.5_{-0.5}^{+0.5}$&$0.090_{-0.001}^{+0.001}$&$0.478_{-0.003}^{+0.003}$&$0.658_{-0.002}^{+0.002}$&$1.02(120)$ \\\hline
8&4202200116&59392.08&986&103.7 &$0.55\pm0.020$&$14.6_{-0.6}^{+0.8}$&$2.63_{-0.14}^{+0.18}$&$1.00_{-0.05}^{}$&$0.18_{-0.07}^{+0.07}$&$55.0_{-20.2}^{+70.5}$&$0.107_{-0.001}^{+0.001}$&$0.512_{-0.003}^{+0.003}$&$0.709_{-0.003}^{+0.003}$&$0.86(114)$ \\\hline
9&4202200119&59396.59&860&150.4 &$0.50\pm0.048$&$17.4_{-1.7}^{+1.5}$&$2.41_{-0.16}^{+0.25}$&$0.69_{-0.20}^{+0.18}$&$0.21_{-0.02}^{+0.05}$&$55.7_{-17.8}^{+30.5}$&$0.218_{-0.001}^{+0.001}$&$0.719_{-0.004}^{+0.004}$&$1.027_{-0.004}^{+0.004}$&$0.81(121)$ \\\hline
10&4202200120&59397.94&165&194.2 &$0.55\pm0.071$&$15.9_{-1.9}^{+1.7}$&$2.57_{-0.30}^{+0.51}$&$0.52_{-0.21}^{+0.20}$&$0.19_{-0.03}^{+0.03}$&$92.5_{-27.8}^{+100.9}$&$0.402_{-0.004}^{+0.003}$&$0.882_{-0.010}^{+0.010}$&$1.375_{-0.010}^{+0.010}$&$0.98(96)$ \\\hline
11&4202200121&59398.20&2146&233.5 &$0.52\pm0.025$&$14.3_{-0.6}^{+0.7}$&$2.95_{-0.17}^{+0.18}$&$0.57_{-0.04}^{+0.04}$&$0.27_{-0.02}^{+0.02}$&$52.9_{-11.1}^{+13.8}$&$0.480_{-0.001}^{+0.001}$&$1.029_{-0.004}^{+0.004}$&$1.599_{-0.004}^{+0.004}$&$0.85(135)$ \\\hline
12&4202200122&59401.69&1555&666.3 &$0.51\pm0.007$&$14.4_{-1.0}^{+1.9}$&$2.16_{-0.14}^{+0.13}$&$0.91_{-0.16}^{}$&$0.61_{-0.06}^{+0.07}$&$17.4_{-2.6}^{+3.8}$&$1.527_{-0.003}^{+0.003}$&$1.979_{-0.008}^{+0.008}$&$3.597_{-0.007}^{+0.007}$&$0.60(140)$ \\\hline
13&4202200123&59402.29&1745&654.2 &$0.51\pm0.007$&$16.3_{-1.3}^{+2.2}$&$2.05_{-0.12}^{+0.12}$&$0.81_{-0.15}^{+0.14}$&$0.62_{-0.07}^{+0.07}$&$16.9_{-2.4}^{+4.4}$&$1.518_{-0.003}^{+0.003}$&$2.004_{-0.007}^{+0.007}$&$3.612_{-0.007}^{+0.007}$&$0.64(142)$ \\\hline
14&4202200124&59403.06&2152&566.2 &$0.51\pm0.007$&$17.1_{-2.2}^{+7.0}$&$1.92_{-0.19}^{+0.07}$&$0.59_{-0.24}^{+0.13}$&$0.70_{-0.05}^{+0.08}$&$13.1_{-1.9}^{+1.7}$&$1.499_{-0.094}^{+0.127}$&$1.275_{-0.095}^{+0.127}$&$2.864_{-0.011}^{+0.006}$&$0.56(141)$ \\\hline
15&4202200126&59405.06&3526&556.9 &$0.51\pm0.006$&$17.7_{-2.0}^{+5.4}$&$1.97_{-0.16}^{+0.12}$&$0.58_{-0.19}^{+0.12}$&$0.71_{-0.04}^{+0.07}$&$12.6_{-1.7}^{+1.5}$&$1.472_{-0.003}^{+0.003}$&$1.373_{-0.006}^{+0.006}$&$2.936_{-0.005}^{+0.005}$&$0.64(147)$ \\\hline
16&4202200127&59406.47&3064&524.0 &$0.51\pm0.006$&$16.4_{-1.6}^{+3.2}$&$2.07_{-0.15}^{+0.13}$&$0.67_{-0.16}^{+0.11}$&$0.68_{-0.05}^{+0.06}$&$12.9_{-1.6}^{+1.6}$&$1.332_{-0.003}^{+0.003}$&$1.406_{-0.006}^{+0.006}$&$2.829_{-0.005}^{+0.005}$&$0.55(146)$ \\\hline
17&4202200128&59407.50&2389&491.3 &$0.51\pm0.007$&$14.3_{-1.4}^{+2.7}$&$2.30_{-0.21}^{+0.18}$&$0.73_{-0.16}^{+0.11}$&$0.67_{-0.04}^{+0.06}$&$13.0_{-1.6}^{+1.7}$&$1.226_{-0.003}^{+0.003}$&$1.347_{-0.006}^{+0.006}$&$2.663_{-0.005}^{+0.005}$&$0.77(142)$ \\\hline
18&4202200129&59408.47&3469&458.6 &$0.51\pm0.006$&$23.6_{-5.1}^{}$&$1.76_{-0.19}^{+0.13}$&$0.34_{-0.22}^{+0.14}$&$0.75_{-0.05}^{+0.11}$&$10.5_{-1.9}^{+1.3}$&$1.308_{-0.003}^{+0.003}$&$0.885_{-0.005}^{+0.005}$&$2.283_{-0.004}^{+0.004}$&$0.56(144)$ \\\hline
19&4202200130&59417.93&790&218.6 &$0.50\pm0.009$&$22.3_{-5.6}^{+13.2}$&$1.83_{-0.20}^{+0.29}$&$0.26_{-0.10}^{+0.10}$&$0.60_{-0.04}^{+0.04}$&$10.8_{-1.1}^{+1.3}$&$0.555_{-0.002}^{+0.002}$&$0.362_{-0.004}^{+0.004}$&$1.007_{-0.004}^{+0.004}$&$0.82(118)$ \\\hline
20&4202200131&59418.05&1927&220.5 &$0.51\pm0.009$&$15.8_{-2.4}^{+3.0}$&$2.16_{-0.21}^{+0.29}$&$0.43_{-0.08}^{+0.08}$&$0.57_{-0.03}^{+0.03}$&$11.9_{-1.0}^{+1.2}$&$0.527_{-0.002}^{+0.002}$&$0.441_{-0.003}^{+0.003}$&$1.057_{-0.003}^{+0.003}$&$0.74(130)$ \\\hline
21&4202200132&59450.90&1003&57.9 &$0.59\pm0.009$&$12.5_{-1.2}^{+1.5}$&$2.67_{-0.27}^{+0.31}$&$0.93_{-0.09}^{}$&$0.11_{-0.01}^{+0.01}$&$101.2_{-0.6}^{+0.6}$&$0.045_{-0.001}^{+0.001}$&$0.232_{-0.002}^{+0.002}$&$0.367_{-0.002}^{+0.002}$&$1.01(111)$ \\\hline
22&4202200133&59451.48&2517&58.9 &$0.58\pm0.008$&$12.9_{-1.3}^{+1.6}$&$2.80_{-0.29}^{+0.35}$&$0.82_{-0.07}^{+0.06}$&$0.11_{-0.01}^{+0.01}$&$103.7_{-0.4}^{+0.4}$&$0.046_{-0.001}^{+0.001}$&$0.242_{-0.001}^{+0.001}$&$0.378_{-0.001}^{+0.001}$&$0.88(121)$ \\\hline
23&4202200134&59455.23&1153&53.4 &$0.58\pm0.056$&$10.7_{-1.3}^{+3.1}$&$2.96_{-0.62}^{+0.31}$&$1.00_{-0.12}^{}$&$0.16_{-0.15}^{+0.07}$&$44.0_{-20.7}^{+103.2}$&$0.038_{-0.001}^{+0.001}$&$0.181_{-0.002}^{+0.002}$&$0.308_{-0.002}^{+0.002}$&$0.92(112)$ \\\hline
24&4202200135&59624.52&6923&602.4 &$0.51\pm0.006$&$18.5_{-2.5}^{+13.8}$&$1.85_{-0.12}^{+0.11}$&$0.50_{-0.27}^{+0.13}$&$0.75_{-0.04}^{+0.09}$&$12.0_{-1.3}^{+1.3}$&$1.667_{-0.003}^{+0.003}$&$1.240_{-0.005}^{+0.005}$&$2.998_{-0.005}^{+0.005}$&$0.58(154)$ \\\hline
25&4202200136&59625.11&3908&645.6 &$0.51\pm0.006$&$16.6_{-1.3}^{+3.5}$&$1.98_{-0.13}^{+0.09}$&$0.74_{-0.20}^{+0.06}$&$0.68_{-0.05}^{+0.08}$&$14.2_{-2.1}^{+1.9}$&$1.605_{-0.003}^{+0.003}$&$1.776_{-0.006}^{+0.006}$&$3.470_{-0.006}^{+0.006}$&$0.55(150)$ \\\hline
26&4202200137&59626.01&1842&645.5 &$0.51\pm0.006$&$24.0_{-6.5}^{}$&$1.74_{-0.16}^{+0.17}$&$0.37_{-0.21}^{+0.24}$&$0.79_{-0.08}^{+0.13}$&$11.4_{-2.2}^{+2.4}$&$1.880_{-0.004}^{+0.004}$&$1.339_{-0.007}^{+0.007}$&$3.309_{-0.006}^{+0.006}$&$0.66(141)$ \\\hline
27&4202200138&59627.05&4099&652.3 &$0.51\pm0.006$&$16.5_{-2.0}^{+8.9}$&$1.98_{-0.21}^{+0.13}$&$0.61_{-0.30}^{+0.14}$&$0.75_{-0.05}^{+0.10}$&$12.5_{-3.1}^{+1.6}$&$1.762_{-0.003}^{+0.003}$&$1.516_{-0.006}^{+0.006}$&$3.368_{-0.005}^{+0.005}$&$0.62(150)$ \\\hline
28&4202200139&59628.01&1110&661.9 &$0.52\pm0.008$&$14.3_{-1.2}^{+2.0}$&$2.06_{-0.15}^{+0.15}$&$0.89_{-0.18}^{}$&$0.59_{-0.07}^{+0.07}$&$18.7_{-2.8}^{+4.9}$&$1.508_{-0.109}^{+0.141}$&$1.817_{-0.116}^{+0.146}$&$3.415_{-0.038}^{+0.036}$&$0.66(137)$ \\\hline
29&4202200140&59629.05&2372&627.8 &$0.51\pm0.007$&$13.0_{-1.2}^{+1.6}$&$2.19_{-0.15}^{+0.18}$&$0.88_{-0.15}^{}$&$0.63_{-0.06}^{+0.05}$&$16.4_{-1.9}^{+3.6}$&$1.496_{-0.096}^{+0.104}$&$1.586_{-0.100}^{+0.106}$&$3.172_{-0.029}^{+0.023}$&$0.73(144)$ \\\hline
30&5202200102&59640.10&7562&589.2 &$0.50\pm0.004$&$20.4_{-3.6}^{}$&$1.79_{-0.19}^{+0.12}$&$0.43_{-0.30}^{+0.15}$&$0.75_{-0.05}^{+0.14}$&$11.8_{-2.3}^{+1.6}$&$1.643_{-0.113}^{+0.269}$&$1.127_{-0.113}^{+0.269}$&$2.859_{-0.009}^{+0.009}$&$0.52(147)$ \\\hline
31&5202200103&59641.33&1982&584.4 &$0.50\pm0.009$&$13.4_{-1.8}^{+3.5}$&$2.17_{-0.24}^{+0.23}$&$0.71_{-0.21}^{+0.15}$&$0.68_{-0.05}^{+0.06}$&$13.9_{-1.7}^{+1.9}$&$1.507_{-0.089}^{+0.090}$&$1.246_{-0.093}^{+0.095}$&$2.842_{-0.028}^{+0.031}$&$0.66(136)$ \\\hline
32&5202200104&59642.04&12902&679.9 &$0.50\pm0.005$&$20.4_{-3.1}^{+43.5}$&$1.86_{-0.27}^{+0.10}$&$0.50_{-0.37}^{+0.15}$&$0.78_{-0.06}^{+0.15}$&$11.7_{-2.5}^{+1.7}$&$1.917_{-0.150}^{+0.334}$&$1.567_{-0.150}^{+0.334}$&$3.574_{-0.011}^{+0.013}$&$0.48(152)$ \\\hline
33&5202200105&59643.01&5594&698.2 &$0.50\pm0.005$&$24.2_{-6.1}^{}$&$1.76_{-0.14}^{+0.14}$&$0.35_{-0.20}^{+0.20}$&$0.82_{-0.07}^{+0.11}$&$11.2_{-1.7}^{+1.8}$&$2.080_{-0.170}^{+0.296}$&$1.381_{-0.170}^{+0.296}$&$3.550_{-0.011}^{+0.011}$&$0.51(148)$ \\\hline
34&5202200106&59647.09&1082&785.3 &$0.50\pm0.007$&$16.1_{-2.7}^{}$&$2.13_{-0.37}^{+0.25}$&$0.72_{-0.57}^{+0.26}$&$0.77_{-0.11}^{+0.29}$&$12.7_{-3.8}^{+4.0}$&$2.128_{-0.272}^{+0.772}$&$2.176_{-0.276}^{+0.777}$&$4.395_{-0.044}^{+0.089}$&$0.92(135)$ \\\hline
35&5202200107&59648.23&2813&731.0 &$0.50\pm0.006$&$20.1_{-3.9}^{}$&$1.80_{-0.21}^{+0.14}$&$0.47_{-0.24}^{+0.20}$&$0.78_{-0.07}^{+0.20}$&$12.3_{-2.9}^{+2.0}$&$2.087_{-0.166}^{+0.620}$&$1.531_{-0.167}^{+0.621}$&$3.708_{-0.014}^{+0.020}$&$0.53(145)$ \\\hline
36&5202200108&59649.07&2867&720.3 &$0.51\pm0.006$&$15.5_{-1.4}^{+3.4}$&$1.98_{-0.15}^{+0.11}$&$0.76_{-0.22}^{+0.14}$&$0.67_{-0.05}^{+0.08}$&$15.6_{-2.3}^{+2.5}$&$1.819_{-0.133}^{+0.141}$&$1.792_{-0.135}^{+0.144}$&$3.701_{-0.024}^{+0.029}$&$0.51(145)$ \\\hline
37&5202200110&59653.77&793&754.4 &$0.52\pm0.008$&$12.1_{-1.7}^{+3.1}$&$2.35_{-0.30}^{+0.36}$&$0.88_{-0.25}^{}$&$0.70_{-0.07}^{+0.08}$&$14.9_{-2.1}^{+3.0}$&$1.938_{-0.124}^{+0.184}$&$1.895_{-0.136}^{+0.192}$&$3.924_{-0.058}^{+0.057}$&$0.86(135)$ \\\hline
38&5202200111&59656.69&1597&743.5 &$0.51\pm0.007$&$17.6_{-2.8}^{}$&$1.90_{-0.28}^{+0.16}$&$0.63_{-0.46}^{+0.21}$&$0.74_{-0.08}^{+0.23}$&$13.3_{-3.3}^{+2.7}$&$2.001_{-0.178}^{+0.646}$&$1.803_{-0.180}^{+0.647}$&$3.894_{-0.024}^{+0.033}$&$0.59(141)$ \\\hline
39&5202200114&59659.02&6348&751.7 &$0.51\pm0.006$&$17.8_{-1.4}^{+4.4}$&$1.93_{-0.12}^{+0.08}$&$0.69_{-0.21}^{+0.11}$&$0.72_{-0.05}^{+0.09}$&$14.0_{-2.1}^{+1.8}$&$1.959_{-0.136}^{+0.168}$&$2.039_{-0.137}^{+0.170}$&$4.088_{-0.018}^{+0.023}$&$0.47(153)$ \\\hline
40&5202200116&59661.01&4471&742.4 &$0.51\pm0.006$&$15.6_{-1.2}^{+3.3}$&$2.03_{-0.15}^{+0.10}$&$0.79_{-0.22}^{+0.13}$&$0.69_{-0.06}^{+0.08}$&$14.8_{-2.3}^{+2.3}$&$1.880_{-0.143}^{+0.150}$&$1.948_{-0.145}^{+0.153}$&$3.918_{-0.023}^{+0.028}$&$0.68(150)$ \\\hline
41&4639010101&59663.59&7446&704.0 &$0.51\pm0.006$&$17.3_{-1.7}^{+6.5}$&$1.94_{-0.16}^{+0.09}$&$0.62_{-0.25}^{+0.12}$&$0.75_{-0.04}^{+0.09}$&$12.9_{-1.9}^{+1.4}$&$1.936_{-0.131}^{+0.161}$&$1.650_{-0.132}^{+0.163}$&$3.675_{-0.015}^{+0.020}$&$0.54(153)$ \\\hline
42&4639010103&59665.01&5134&638.6 &$0.51\pm0.006$&$12.4_{-1.7}^{+2.9}$&$2.42_{-0.28}^{+0.28}$&$0.68_{-0.17}^{+0.12}$&$0.73_{-0.04}^{+0.05}$&$13.0_{-1.3}^{+1.3}$&$1.735_{-0.073}^{+0.082}$&$1.372_{-0.081}^{+0.091}$&$3.197_{-0.037}^{+0.039}$&$0.54(149)$ \\\hline
43&4639010105&59667.21&1529&719.9 &$0.51\pm0.008$&$17.6_{-1.7}^{+3.7}$&$1.85_{-0.12}^{+0.10}$&$0.76_{-0.21}^{+0.16}$&$0.66_{-0.08}^{+0.08}$&$15.9_{-2.4}^{+3.9}$&$1.745_{-0.166}^{+0.174}$&$1.965_{-0.167}^{+0.177}$&$3.800_{-0.024}^{+0.030}$&$0.76(139)$ \\\hline
44&4639010106&59668.16&3078&746.1 &$0.51\pm0.006$&$16.9_{-1.7}^{+7.5}$&$1.94_{-0.18}^{+0.11}$&$0.71_{-0.32}^{+0.15}$&$0.73_{-0.06}^{+0.12}$&$13.8_{-2.6}^{+2.2}$&$1.990_{-0.185}^{+0.228}$&$1.895_{-0.186}^{+0.230}$&$3.975_{-0.020}^{+0.028}$&$0.66(149)$ \\\hline
45&4639010107&59669.52&4371&757.2 &$0.51\pm0.005$&$17.2_{-1.2}^{+2.2}$&$1.88_{-0.08}^{+0.07}$&$0.79_{-0.15}^{+0.13}$&$0.64_{-0.06}^{+0.06}$&$17.3_{-2.3}^{+3.7}$&$1.792_{-0.141}^{+0.141}$&$2.138_{-0.142}^{+0.143}$&$4.020_{-0.021}^{+0.026}$&$0.58(152)$ \\\hline
46&4639010108&59670.63&4969&720.9 &$0.51\pm0.006$&$15.1_{-1.5}^{+3.8}$&$2.02_{-0.17}^{+0.12}$&$0.72_{-0.22}^{+0.14}$&$0.70_{-0.05}^{+0.07}$&$14.4_{-2.0}^{+2.1}$&$1.855_{-0.125}^{+0.137}$&$1.651_{-0.127}^{+0.140}$&$3.596_{-0.022}^{+0.026}$&$0.67(151)$ \\\hline
47&4639010109&59671.65&3391&718.2 &$0.51\pm0.005$&$20.6_{-3.5}^{}$&$1.79_{-0.20}^{+0.12}$&$0.52_{-0.38}^{+0.19}$&$0.77_{-0.07}^{+0.16}$&$12.5_{-2.8}^{+2.1}$&$1.996_{-0.179}^{+0.379}$&$1.643_{-0.179}^{+0.380}$&$3.729_{-0.013}^{+0.020}$&$0.69(149)$ \\\hline
48&4639010110&59672.70&2900&728.1 &$0.51\pm0.006$&$13.5_{-1.0}^{+2.2}$&$2.38_{-0.21}^{+0.16}$&$0.87_{-0.19}^{+0.12}$&$0.68_{-0.06}^{+0.08}$&$15.0_{-2.2}^{+2.4}$&$1.802_{-0.131}^{+0.132}$&$2.119_{-0.137}^{+0.139}$&$4.011_{-0.043}^{+0.044}$&$0.57(147)$ \\\hline
49&4639010111&59673.58&3561&660.2 &$0.51\pm0.006$&$16.3_{-1.7}^{+4.5}$&$1.94_{-0.16}^{+0.11}$&$0.64_{-0.21}^{+0.13}$&$0.69_{-0.04}^{+0.07}$&$14.4_{-1.9}^{+1.8}$&$1.734_{-0.102}^{+0.124}$&$1.497_{-0.104}^{+0.126}$&$3.321_{-0.020}^{+0.024}$&$0.51(148)$ \\\hline
50&5202200118&59676.11&583&705.7 &$0.51\pm0.007$&$15.7_{-1.8}^{+6.2}$&$1.96_{-0.22}^{+0.17}$&$0.84_{-0.35}^{}$&$0.66_{-0.09}^{+0.13}$&$15.8_{-3.3}^{+4.0}$&$1.707_{-0.191}^{+0.263}$&$1.923_{-0.195}^{+0.267}$&$3.721_{-0.039}^{+0.047}$&$0.80(132)$ \\\hline
51&4639010114&59676.55&4910&707.7 &$0.50\pm0.006$&$13.9_{-1.3}^{+3.4}$&$2.41_{-0.24}^{+0.18}$&$0.76_{-0.21}^{+0.12}$&$0.77_{-0.05}^{+0.09}$&$12.1_{-1.7}^{+1.5}$&$1.912_{-0.122}^{+0.138}$&$1.970_{-0.128}^{+0.146}$&$3.972_{-0.040}^{+0.048}$&$0.58(153)$ \\\hline
52&4639010117&59679.73&3162&680.7 &$0.51\pm0.006$&$16.4_{-1.8}^{+4.5}$&$2.00_{-0.16}^{+0.13}$&$0.65_{-0.20}^{+0.13}$&$0.71_{-0.05}^{+0.07}$&$13.2_{-1.7}^{+1.7}$&$1.619_{-0.099}^{+0.117}$&$1.489_{-0.102}^{+0.120}$&$3.199_{-0.022}^{+0.027}$&$0.62(146)$ \\\hline
53&4639010118&59680.56&4600&596.1 &$0.51\pm0.006$&$14.4_{-1.2}^{+1.7}$&$2.13_{-0.13}^{+0.13}$&$0.78_{-0.13}^{+0.12}$&$0.65_{-0.05}^{+0.05}$&$14.9_{-1.6}^{+2.4}$&$1.458_{-0.085}^{+0.082}$&$1.540_{-0.088}^{+0.087}$&$3.089_{-0.024}^{+0.027}$&$0.46(150)$ \\\hline
54&4639010120&59682.42&4339&641.5 &$0.51\pm0.006$&$19.3_{-3.5}^{+11.6}$&$1.95_{-0.16}^{+0.16}$&$0.50_{-0.24}^{+0.19}$&$0.80_{-0.07}^{+0.07}$&$11.1_{-1.7}^{+1.9}$&$1.807_{-0.129}^{+0.227}$&$1.551_{-0.130}^{+0.229}$&$3.448_{-0.017}^{+0.028}$&$0.53(151)$ \\\hline
55&4639010121&59683.28&4559&638.9 &$0.51\pm0.006$&$17.9_{-1.8}^{+5.4}$&$1.88_{-0.13}^{+0.09}$&$0.61_{-0.21}^{+0.12}$&$0.72_{-0.04}^{+0.08}$&$13.1_{-1.8}^{+1.5}$&$1.698_{-0.100}^{+0.138}$&$1.529_{-0.101}^{+0.139}$&$3.317_{-0.015}^{+0.018}$&$0.61(151)$ \\\hline
56&4639010122&59684.05&3881&595.5 &$0.51\pm0.006$&$13.6_{-1.8}^{+3.2}$&$2.22_{-0.23}^{+0.23}$&$0.64_{-0.17}^{+0.12}$&$0.71_{-0.04}^{+0.05}$&$13.1_{-1.3}^{+1.4}$&$1.601_{-0.069}^{+0.081}$&$1.258_{-0.074}^{+0.086}$&$2.949_{-0.027}^{+0.029}$&$0.56(148)$ \\\hline
57&4639010123&59685.01&1927&673.2 &$0.51\pm0.007$&$14.0_{-1.1}^{+1.4}$&$2.24_{-0.20}^{+0.15}$&$0.89_{-0.21}^{}$&$0.65_{-0.06}^{+0.09}$&$15.8_{-2.7}^{+3.2}$&$1.624_{-0.139}^{+0.138}$&$1.951_{-0.144}^{+0.145}$&$3.665_{-0.038}^{+0.042}$&$0.58(143)$ \\\hline
58&4639010124&59686.63&179&653.2 &$0.51\pm0.009$&$27.7_{-13.5}^{}$&$1.73_{-0.16}^{+0.46}$&$0.28_{-0.17}^{+0.42}$&$0.83_{-0.13}^{+0.11}$&$10.9_{-1.8}^{+2.7}$&$1.990_{-0.273}^{+0.263}$&$1.235_{-0.276}^{+0.278}$&$3.314_{-0.040}^{+0.091}$&$0.83(114)$ \\\hline
59&4639010125&59687.85&507&653.5 &$0.51\pm0.008$&$16.9_{-2.0}^{+3.7}$&$1.90_{-0.16}^{+0.16}$&$0.79_{-0.22}^{+0.20}$&$0.62_{-0.09}^{+0.08}$&$16.7_{-2.7}^{+5.6}$&$1.521_{-0.170}^{+0.167}$&$1.856_{-0.174}^{+0.172}$&$3.468_{-0.039}^{+0.044}$&$0.83(131)$ \\\hline
60&4639010128&59691.03&2047&625.7 &$0.51\pm0.006$&$20.1_{-4.1}^{}$&$1.87_{-0.23}^{+0.17}$&$0.48_{-0.38}^{+0.20}$&$0.79_{-0.07}^{+0.18}$&$11.3_{-2.6}^{+2.0}$&$1.787_{-0.147}^{+0.348}$&$1.415_{-0.148}^{+0.349}$&$3.292_{-0.017}^{+0.028}$&$0.73(142)$ \\\hline
61&4639010129&59692.05&1393&631.5 &$0.50\pm0.007$&$17.1_{-2.8}^{}$&$1.96_{-0.30}^{+0.19}$&$0.66_{-0.53}^{+0.23}$&$0.73_{-0.09}^{+0.21}$&$12.5_{-3.3}^{+3.3}$&$1.666_{-0.179}^{+0.488}$&$1.591_{-0.180}^{+0.489}$&$3.347_{-0.025}^{+0.036}$&$0.69(138)$ \\\hline
62&4639010130&59694.05&3235&542.8 &$0.51\pm0.006$&$13.8_{-1.9}^{+3.5}$&$2.18_{-0.23}^{+0.24}$&$0.59_{-0.16}^{+0.13}$&$0.70_{-0.04}^{+0.05}$&$12.9_{-1.3}^{+1.3}$&$1.475_{-0.063}^{+0.069}$&$1.086_{-0.067}^{+0.075}$&$2.651_{-0.025}^{+0.030}$&$0.77(144)$ \\\hline
63&4639010132&59696.05&1896&554.9 &$0.51\pm0.007$&$15.6_{-2.4}^{+6.0}$&$1.97_{-0.21}^{+0.20}$&$0.56_{-0.21}^{+0.15}$&$0.70_{-0.04}^{+0.06}$&$13.1_{-1.6}^{+1.6}$&$1.506_{-0.076}^{+0.106}$&$1.094_{-0.079}^{+0.109}$&$2.690_{-0.020}^{+0.028}$&$0.74(140)$ \\\hline
64&4639010133&59698.37&1799&635.5 &$0.51\pm0.007$&$15.7_{-1.8}^{+4.0}$&$1.98_{-0.16}^{+0.15}$&$0.70_{-0.21}^{+0.15}$&$0.67_{-0.05}^{+0.07}$&$14.7_{-2.0}^{+2.3}$&$1.621_{-0.110}^{+0.126}$&$1.534_{-0.113}^{+0.130}$&$3.244_{-0.025}^{+0.032}$&$0.61(141)$ \\\hline
65&4639010134&59700.05&450&598.4 &$0.52\pm0.007$&$24.3_{-9.4}^{}$&$1.68_{-0.17}^{+0.31}$&$0.30_{-0.20}^{+0.30}$&$0.77_{-0.08}^{+0.11}$&$11.6_{-1.8}^{+2.5}$&$1.778_{-0.169}^{+0.230}$&$1.001_{-0.170}^{+0.233}$&$2.869_{-0.021}^{+0.042}$&$0.66(125)$ \\\hline
66&4639010135&59702.11&2992&704.9 &$0.51\pm0.006$&$15.8_{-1.3}^{+3.1}$&$2.09_{-0.15}^{+0.11}$&$0.78_{-0.20}^{+0.12}$&$0.70_{-0.06}^{+0.08}$&$14.4_{-2.1}^{+2.2}$&$1.789_{-0.129}^{+0.146}$&$1.999_{-0.132}^{+0.149}$&$3.878_{-0.028}^{+0.031}$&$0.56(149)$ \\\hline
67&4639010136&59703.21&2215&628.6 &$0.51\pm0.006$&$17.4_{-2.1}^{+8.6}$&$1.94_{-0.19}^{+0.13}$&$0.65_{-0.30}^{+0.15}$&$0.73_{-0.06}^{+0.11}$&$12.8_{-2.2}^{+2.0}$&$1.716_{-0.136}^{+0.200}$&$1.586_{-0.138}^{+0.202}$&$3.393_{-0.020}^{+0.027}$&$0.54(144)$ \\\hline
68&4639010137&59704.05&2446&702.2 &$0.51\pm0.006$&$17.0_{-1.6}^{+6.4}$&$2.00_{-0.18}^{+0.10}$&$0.73_{-0.29}^{+0.13}$&$0.71_{-0.06}^{+0.12}$&$13.7_{-2.2}^{+2.1}$&$1.837_{-0.161}^{+0.203}$&$1.929_{-0.163}^{+0.205}$&$3.857_{-0.026}^{+0.031}$&$0.66(147)$ \\\hline
69&4639010138&59705.08&4848&630.9 &$0.51\pm0.006$&$15.6_{-1.8}^{+4.3}$&$1.99_{-0.16}^{+0.13}$&$0.62_{-0.19}^{+0.13}$&$0.72_{-0.04}^{+0.06}$&$13.2_{-1.6}^{+1.5}$&$1.704_{-0.093}^{+0.109}$&$1.347_{-0.095}^{+0.110}$&$3.141_{-0.021}^{+0.019}$&$0.58(152)$ \\\hline
70&4639010139&59706.04&2998&631.1 &$0.51\pm0.006$&$17.0_{-2.1}^{+6.2}$&$1.98_{-0.18}^{+0.13}$&$0.58_{-0.22}^{+0.13}$&$0.73_{-0.04}^{+0.08}$&$12.8_{-1.7}^{+1.5}$&$1.726_{-0.100}^{+0.137}$&$1.446_{-0.102}^{+0.140}$&$3.261_{-0.020}^{+0.026}$&$0.65(147)$ \\\hline
71&4639010140&59715.35&5811&477.8 &$0.51\pm0.006$&$13.8_{-1.9}^{+2.2}$&$2.37_{-0.20}^{+0.27}$&$0.50_{-0.09}^{+0.09}$&$0.69_{-0.03}^{+0.03}$&$12.6_{-0.8}^{+1.0}$&$1.301_{-0.032}^{+0.039}$&$0.963_{-0.040}^{+0.049}$&$2.353_{-0.025}^{+0.030}$&$0.54(149)$ \\\hline
72&4639010141&59718.12&4741&581.1 &$0.51\pm0.006$&$17.7_{-2.0}^{+6.1}$&$1.97_{-0.16}^{+0.11}$&$0.59_{-0.21}^{+0.12}$&$0.74_{-0.05}^{+0.08}$&$12.0_{-1.7}^{+1.4}$&$1.576_{-0.093}^{+0.129}$&$1.406_{-0.095}^{+0.131}$&$3.072_{-0.018}^{+0.020}$&$0.53(151)$ \\\hline
73&4639010143&59719.99&3886&583.8 &$0.51\pm0.006$&$18.7_{-2.3}^{+10.1}$&$1.89_{-0.17}^{+0.11}$&$0.53_{-0.24}^{+0.12}$&$0.73_{-0.04}^{+0.09}$&$12.2_{-1.8}^{+1.3}$&$1.604_{-0.099}^{+0.154}$&$1.325_{-0.100}^{+0.154}$&$3.019_{-0.015}^{+0.014}$&$0.58(148)$ \\\hline
74&4639010144&59721.05&5045&614.2 &$0.51\pm0.007$&$14.6_{-1.3}^{+2.7}$&$2.17_{-0.17}^{+0.13}$&$0.73_{-0.17}^{+0.11}$&$0.70_{-0.04}^{+0.06}$&$13.4_{-1.6}^{+1.6}$&$1.592_{-0.089}^{+0.096}$&$1.557_{-0.092}^{+0.100}$&$3.240_{-0.025}^{+0.028}$&$0.58(152)$ \\\hline
75&4639010145&59722.33&6456&523.8 &$0.50\pm0.006$&$15.5_{-1.9}^{+4.0}$&$2.04_{-0.17}^{+0.15}$&$0.52_{-0.15}^{+0.11}$&$0.72_{-0.03}^{+0.05}$&$12.2_{-1.1}^{+1.0}$&$1.441_{-0.055}^{+0.067}$&$1.014_{-0.058}^{+0.070}$&$2.545_{-0.017}^{+0.019}$&$0.66(151)$ \\\hline
76&4639010147&59724.01&7083&565.7 &$0.51\pm0.006$&$16.0_{-2.2}^{+4.3}$&$2.00_{-0.16}^{+0.16}$&$0.51_{-0.16}^{+0.12}$&$0.73_{-0.04}^{+0.05}$&$12.2_{-1.1}^{+1.2}$&$1.573_{-0.064}^{+0.083}$&$1.082_{-0.066}^{+0.085}$&$2.745_{-0.015}^{+0.020}$&$0.67(153)$ \\\hline
77&4639010148&59725.10&302&603.0 &$0.51\pm0.008$&$30.5_{-15.1}^{}$&$1.62_{-0.14}^{+0.18}$&$0.24_{-0.13}^{+0.39}$&$0.80_{-0.12}^{+0.09}$&$11.1_{-3.0}^{+3.1}$&$1.854_{-0.271}^{+0.455}$&$0.995_{-0.272}^{+0.456}$&$2.938_{-0.026}^{+0.032}$&$0.82(119)$ \\\hline
78&4639010149&59728.83&1705&652.6 &$0.51\pm0.006$&$13.9_{-1.5}^{+3.6}$&$2.32_{-0.25}^{+0.22}$&$0.78_{-0.23}^{+0.14}$&$0.72_{-0.06}^{+0.08}$&$13.2_{-2.0}^{+2.0}$&$1.708_{-0.120}^{+0.142}$&$1.799_{-0.128}^{+0.151}$&$3.597_{-0.043}^{+0.049}$&$0.64(142)$ \\\hline
79&4639010150&59729.09&4382&612.0 &$0.51\pm0.004$&$17.7_{-2.0}^{+8.1}$&$1.92_{-0.18}^{+0.11}$&$0.55_{-0.23}^{+0.11}$&$0.73_{-0.04}^{+0.08}$&$12.8_{-1.9}^{+1.3}$&$1.683_{-0.094}^{+0.140}$&$1.339_{-0.096}^{+0.141}$&$3.112_{-0.019}^{+0.015}$&$0.59(150)$ \\\hline
80&4639010151&59730.06&2401&578.5 &$0.51\pm0.007$&$12.6_{-1.4}^{+1.9}$&$2.32_{-0.21}^{+0.24}$&$0.81_{-0.15}^{+0.14}$&$0.65_{-0.05}^{+0.05}$&$14.8_{-1.6}^{+2.3}$&$1.437_{-0.079}^{+0.077}$&$1.418_{-0.085}^{+0.085}$&$2.945_{-0.032}^{+0.035}$&$0.61(143)$ \\\hline
81&4639010153&59732.96&1963&586.6 &$0.51\pm0.007$&$13.2_{-1.3}^{+1.6}$&$2.35_{-0.18}^{+0.22}$&$0.87_{-0.15}^{}$&$0.65_{-0.07}^{+0.06}$&$14.7_{-1.8}^{+3.2}$&$1.403_{-0.098}^{+0.100}$&$1.675_{-0.104}^{+0.108}$&$3.168_{-0.035}^{+0.041}$&$0.63(142)$ \\\hline
82&4639010154&59733.09&2840&584.9 &$0.51\pm0.006$&$18.4_{-2.1}^{+5.9}$&$1.90_{-0.14}^{+0.11}$&$0.59_{-0.20}^{+0.12}$&$0.71_{-0.05}^{+0.08}$&$12.9_{-1.8}^{+1.7}$&$1.543_{-0.098}^{+0.129}$&$1.453_{-0.100}^{+0.131}$&$3.085_{-0.018}^{+0.022}$&$0.55(146)$ \\\hline
83&4639010156&59735.89&543&473.2 &$0.51\pm0.008$&$16.3_{-3.9}^{+14.8}$&$1.97_{-0.32}^{+0.38}$&$0.48_{-0.27}^{+0.19}$&$0.68_{-0.05}^{+0.09}$&$12.7_{-2.0}^{+2.0}$&$1.290_{-0.079}^{+0.135}$&$0.890_{-0.083}^{+0.142}$&$2.270_{-0.027}^{+0.045}$&$0.70(124)$ \\\hline
84&4639010157&59736.02&2871&461.8 &$0.50\pm0.006$&$18.6_{-3.4}^{+13.0}$&$1.95_{-0.22}^{+0.19}$&$0.38_{-0.19}^{+0.12}$&$0.73_{-0.04}^{+0.07}$&$11.1_{-1.4}^{+1.0}$&$1.309_{-0.057}^{+0.101}$&$0.823_{-0.058}^{+0.104}$&$2.222_{-0.013}^{+0.023}$&$0.75(141)$ \\\hline
85&4639010158&59737.09&3290&437.7 &$0.51\pm0.004$&$17.4_{-2.8}^{+6.9}$&$2.05_{-0.21}^{+0.20}$&$0.38_{-0.14}^{+0.10}$&$0.72_{-0.03}^{+0.05}$&$11.2_{-1.0}^{+0.9}$&$1.239_{-0.044}^{+0.059}$&$0.786_{-0.047}^{+0.064}$&$2.115_{-0.017}^{+0.025}$&$0.88(143)$ \\\hline
86&4639010159&59738.52&5003&451.8 &$0.51\pm0.005$&$16.5_{-2.1}^{+3.0}$&$1.98_{-0.13}^{+0.15}$&$0.53_{-0.11}^{+0.11}$&$0.67_{-0.04}^{+0.04}$&$12.5_{-1.0}^{+1.3}$&$1.179_{-0.044}^{+0.051}$&$0.943_{-0.047}^{+0.055}$&$2.212_{-0.016}^{+0.019}$&$0.48(147)$ \\\hline
87&4639010161&59740.59&2486&418.8 &$0.51\pm0.006$&$18.0_{-3.4}^{+6.6}$&$2.01_{-0.19}^{+0.23}$&$0.35_{-0.12}^{+0.11}$&$0.70_{-0.03}^{+0.04}$&$11.5_{-1.0}^{+1.0}$&$1.175_{-0.040}^{+0.054}$&$0.731_{-0.044}^{+0.060}$&$1.996_{-0.017}^{+0.025}$&$0.70(140)$ \\\hline
88&4639010162&59741.50&2384&416.7 &$0.51\pm0.005$&$16.1_{-2.7}^{+4.6}$&$2.09_{-0.21}^{+0.25}$&$0.42_{-0.12}^{+0.11}$&$0.68_{-0.03}^{+0.04}$&$12.0_{-1.0}^{+1.1}$&$1.137_{-0.034}^{+0.052}$&$0.764_{-0.039}^{+0.058}$&$1.991_{-0.020}^{+0.026}$&$0.74(140)$ \\\hline
89&4639010164&59743.35&1795&497.3 &$0.51\pm0.006$&$22.1_{-4.8}^{}$&$1.81_{-0.21}^{+0.16}$&$0.37_{-0.22}^{+0.16}$&$0.75_{-0.05}^{+0.13}$&$10.9_{-2.1}^{+1.4}$&$1.425_{-0.096}^{+0.208}$&$0.974_{-0.097}^{+0.209}$&$2.489_{-0.013}^{+0.020}$&$0.76(139)$ \\\hline
90&4639010165&59745.49&1759&467.5 &$0.52\pm0.007$&$8.5_{-1.8}^{+1.9}$&$4.33_{-0.97}^{+1.82}$&$1.00_{-0.12}^{}$&$0.58_{-0.03}^{+0.06}$&$14.5_{-1.9}^{+1.3}$&$0.846_{-0.026}^{+0.051}$&$1.429_{-0.094}^{+0.145}$&$2.364_{-0.090}^{+0.135}$&$0.70(136)$ \\\hline
91&4639010167&59748.06&3385&432.3 &$0.50\pm0.006$&$24.5_{-7.0}^{+16.2}$&$1.74_{-0.12}^{+0.20}$&$0.28_{-0.12}^{+0.17}$&$0.74_{-0.06}^{+0.05}$&$10.6_{-1.0}^{+1.5}$&$1.219_{-0.055}^{+0.092}$&$0.752_{-0.056}^{+0.093}$&$2.061_{-0.011}^{+0.016}$&$0.64(143)$ \\\hline
92&4639010168&59749.09&1933&452.4 &$0.51\pm0.007$&$15.1_{-2.1}^{+3.3}$&$2.07_{-0.18}^{+0.21}$&$0.63_{-0.15}^{+0.14}$&$0.67_{-0.05}^{+0.05}$&$12.6_{-1.3}^{+1.8}$&$1.151_{-0.058}^{+0.066}$&$1.023_{-0.062}^{+0.071}$&$2.265_{-0.023}^{+0.028}$&$0.63(138)$ \\\hline
93&4639010169&59750.06&677&439.4 &$0.51\pm0.010$&$11.1_{-2.5}^{+2.6}$&$2.66_{-0.43}^{+0.87}$&$0.80_{-0.16}^{+0.20}$&$0.60_{-0.07}^{+0.05}$&$15.0_{-1.8}^{+3.8}$&$1.057_{-0.070}^{+0.062}$&$1.082_{-0.088}^{+0.097}$&$2.229_{-0.053}^{+0.075}$&$0.82(127)$ \\\hline
94&4639010170&59751.22&4721&380.8 &$0.50\pm0.006$&$17.9_{-2.9}^{+4.2}$&$2.04_{-0.16}^{+0.20}$&$0.35_{-0.08}^{+0.08}$&$0.67_{-0.03}^{+0.03}$&$11.7_{-0.8}^{+0.9}$&$1.040_{-0.002}^{+0.002}$&$0.665_{-0.004}^{+0.004}$&$1.795_{-0.003}^{+0.003}$&$0.81(144)$ \\\hline
95&4639010171&59752.45&1351&391.0 &$0.50\pm0.008$&$15.8_{-3.6}^{+7.5}$&$2.16_{-0.30}^{+0.41}$&$0.42_{-0.17}^{+0.14}$&$0.69_{-0.04}^{+0.03}$&$11.2_{-1.2}^{+1.3}$&$1.059_{-0.003}^{+0.003}$&$0.715_{-0.005}^{+0.005}$&$1.864_{-0.004}^{+0.004}$&$1.00(134)$ \\\hline
96&4639010172&59753.68&3837&376.9 &$0.50\pm0.006$&$20.1_{-3.9}^{+8.9}$&$1.97_{-0.17}^{+0.20}$&$0.31_{-0.11}^{+0.10}$&$0.72_{-0.03}^{+0.04}$&$10.4_{-0.9}^{+0.9}$&$1.062_{-0.002}^{+0.002}$&$0.651_{-0.004}^{+0.004}$&$1.803_{-0.003}^{+0.003}$&$0.72(142)$ \\\hline
97&4639010173&59754.40&4201&371.4 &$0.51\pm0.005$&$15.6_{-2.0}^{+2.6}$&$2.09_{-0.15}^{+0.19}$&$0.50_{-0.09}^{+0.09}$&$0.65_{-0.03}^{+0.03}$&$12.0_{-0.9}^{+1.1}$&$0.961_{-0.002}^{+0.002}$&$0.742_{-0.004}^{+0.004}$&$1.794_{-0.003}^{+0.003}$&$0.68(143)$ \\\hline
98&4639010174&59755.44&507&377.5 &$0.50\pm0.008$&$23.9_{-8.3}^{}$&$1.65_{-0.19}^{+0.20}$&$0.38_{-0.25}^{+0.19}$&$0.67_{-0.06}^{+0.11}$&$11.7_{-2.5}^{+2.3}$&$0.986_{-0.003}^{+0.003}$&$0.760_{-0.007}^{+0.007}$&$1.836_{-0.006}^{+0.006}$&$0.72(120)$ \\\hline
99&4639010176&59757.50&1513&301.1 &$0.51\pm0.009$&$12.4_{-2.4}^{+2.8}$&$2.56_{-0.37}^{+0.61}$&$0.55_{-0.11}^{+0.10}$&$0.60_{-0.03}^{+0.03}$&$12.7_{-1.1}^{+1.4}$&$0.749_{-0.002}^{+0.002}$&$0.630_{-0.004}^{+0.004}$&$1.469_{-0.004}^{+0.004}$&$0.75(131)$ \\\hline
100&4639010177&59760.00&3048&273.4 &$0.50\pm0.007$&$19.0_{-3.1}^{+5.2}$&$1.91_{-0.16}^{+0.19}$&$0.34_{-0.09}^{+0.08}$&$0.63_{-0.03}^{+0.03}$&$11.1_{-0.9}^{+1.0}$&$0.699_{-0.002}^{+0.002}$&$0.476_{-0.003}^{+0.003}$&$1.265_{-0.003}^{+0.003}$&$0.88(138)$ \\\hline
101&4639010178&59761.03&3847&265.2 &$0.50\pm0.005$&$18.4_{-2.5}^{+3.6}$&$1.92_{-0.13}^{+0.15}$&$0.36_{-0.07}^{+0.07}$&$0.61_{-0.02}^{+0.03}$&$11.6_{-0.8}^{+0.9}$&$0.671_{-0.002}^{+0.002}$&$0.465_{-0.003}^{+0.003}$&$1.226_{-0.003}^{+0.003}$&$0.84(140)$ \\\hline
102&4639010180&59763.03&1819&283.6 &$0.51\pm0.007$&$24.3_{-5.4}^{+18.4}$&$1.72_{-0.15}^{+0.17}$&$0.29_{-0.13}^{+0.11}$&$0.66_{-0.04}^{+0.03}$&$10.3_{-1.2}^{+1.2}$&$0.740_{-0.002}^{+0.002}$&$0.499_{-0.004}^{+0.004}$&$1.329_{-0.003}^{+0.003}$&$0.80(134)$ \\\hline
103&4639010181&59764.00&3600&254.2 &$0.50\pm0.006$&$11.9_{-2.0}^{+2.1}$&$2.87_{-0.39}^{+0.63}$&$0.53_{-0.07}^{+0.07}$&$0.58_{-0.02}^{+0.02}$&$12.0_{-0.8}^{+1.0}$&$0.612_{-0.002}^{+0.002}$&$0.568_{-0.003}^{+0.003}$&$1.270_{-0.003}^{+0.003}$&$0.89(139)$ \\\hline
104&4639010182&59765.16&1409&235.9 &$0.51\pm0.009$&$16.9_{-2.8}^{+4.0}$&$1.99_{-0.20}^{+0.27}$&$0.39_{-0.09}^{+0.08}$&$0.56_{-0.03}^{+0.03}$&$12.7_{-1.1}^{+1.3}$&$0.578_{-0.002}^{+0.002}$&$0.426_{-0.003}^{+0.003}$&$1.094_{-0.003}^{+0.003}$&$0.82(127)$ \\\hline
105&4639010184&59767.22&972&177.4 &$0.47\pm0.010$&$19.2_{-3.7}^{+5.5}$&$1.91_{-0.20}^{+0.28}$&$0.29_{-0.07}^{+0.07}$&$0.52_{-0.03}^{+0.03}$&$12.5_{-1.1}^{+1.4}$&$0.406_{-0.002}^{+0.002}$&$0.287_{-0.003}^{+0.003}$&$0.782_{-0.003}^{+0.003}$&$0.92(118)$ \\\hline
106&4639010185&59768.02&2603&88.6 &$0.45\pm0.021$&$18.5_{-2.5}^{+2.3}$&$2.46_{-0.29}^{+0.52}$&$0.35_{-0.04}^{+0.04}$&$0.29_{-0.02}^{+0.02}$&$24.3_{-3.8}^{+5.9}$&$0.138_{-0.001}^{+0.001}$&$0.266_{-0.002}^{+0.002}$&$0.495_{-0.002}^{+0.002}$&$0.89(99)$ \\\hline
107&4639010186&59769.44&2875&62.5 &$0.43\pm0.025$&$16.9_{-2.0}^{+3.0}$&$2.22_{-0.23}^{+0.26}$&$0.60_{-0.12}^{+0.08}$&$0.26_{-0.04}^{+0.05}$&$19.2_{-5.4}^{+9.9}$&$0.053_{-0.001}^{+0.001}$&$0.186_{-0.001}^{+0.001}$&$0.329_{-0.001}^{+0.001}$&$0.88(124)$ \\\hline
108&4639010187&59773.69&1057&56.0 &$0.53\pm0.007$&$10.6_{-1.2}^{+1.3}$&$3.49_{-0.43}^{+0.58}$&$0.85_{-0.03}^{+0.06}$&$0.09_{-0.01}^{+0.01}$&$140.4_{-0.9}^{+0.9}$&$0.043_{-0.001}^{+0.001}$&$0.234_{-0.002}^{+0.002}$&$0.367_{-0.002}^{+0.002}$&$1.26(110)$ \\\hline
109&4639010188&59774.34&1450&55.1 &$0.50\pm0.006$&$12.6_{-0.9}^{+1.1}$&$2.77_{-0.21}^{+0.24}$&$0.81_{-0.06}^{+0.05}$&$0.08_{-0.01}^{+0.01}$&$176.0_{-1.1}^{+1.1}$&$0.037_{-0.001}^{+0.001}$&$0.207_{-0.001}^{+0.001}$&$0.334_{-0.001}^{+0.001}$&$1.08(115)$ \\\hline
110&4639010189&59781.70&2810&60.5 &$0.54\pm0.006$&$6.7_{-2.2}^{+1.5}$&$8.05_{-2.16}^{+6.77}$&$0.85_{-0.04}^{+0.06}$&$0.10_{-0.01}^{+0.01}$&$132.9_{-0.5}^{+0.5}$&$0.053_{-0.001}^{+0.001}$&$0.389_{-0.002}^{+0.002}$&$0.532_{-0.002}^{+0.002}$&$0.79(125)$ \\\hline
111&4639010190&59782.02&5429&59.9 &$0.53\pm0.005$&$7.1_{-1.6}^{+1.1}$&$7.20_{-1.77}^{+3.29}$&$0.86_{-0.03}^{+0.04}$&$0.10_{-0.01}^{+0.01}$&$129.5_{-0.4}^{+0.4}$&$0.051_{-0.001}^{+0.001}$&$0.361_{-0.001}^{+0.001}$&$0.502_{-0.001}^{+0.001}$&$0.79(133)$ \\\hline
112&4639010191&59783.05&6267&60.0 &$0.52\pm0.005$&$7.1_{-2.2}^{+1.3}$&$8.51_{-1.99}^{+6.50}$&$0.74_{-0.02}^{+0.04}$&$0.11_{-0.01}^{+0.01}$&$122.7_{-0.3}^{+0.3}$&$0.053_{-0.001}^{+0.001}$&$0.438_{-0.001}^{+0.001}$&$0.581_{-0.001}^{+0.001}$&$0.96(135)$ \\\hline
113&4639010192&59784.02&3321&57.6 &$0.54\pm0.005$&$9.7_{-1.2}^{+1.2}$&$4.33_{-0.62}^{+0.89}$&$0.81_{-0.04}^{+0.04}$&$0.09_{-0.01}^{+0.01}$&$149.5_{-0.6}^{+0.6}$&$0.047_{-0.001}^{+0.001}$&$0.278_{-0.001}^{+0.001}$&$0.415_{-0.001}^{+0.001}$&$0.78(125)$ \\\hline
114&4639010193&59785.18&479&53.7 &$0.57\pm0.009$&$7.6_{-1.5}^{+1.2}$&$4.45_{-0.96}^{+1.93}$&$0.98_{-0.08}^{}$&$0.09_{-0.01}^{+0.01}$&$168.4_{-1.8}^{+1.8}$&$0.041_{-0.001}^{+0.001}$&$0.194_{-0.003}^{+0.003}$&$0.325_{-0.002}^{+0.002}$&$0.94(99)$ \\\hline
115&4639010194&59786.35&657&48.5 &$0.57\pm0.008$&$4.5_{-2.6}^{+2.2}$&$12.05_{-5.34}^{+35.92}$&$0.87_{-0.08}^{}$&$0.10_{-0.01}^{+0.01}$&$115.0_{-1.0}^{+1.1}$&$0.038_{-0.001}^{+0.001}$&$0.279_{-0.003}^{+0.003}$&$0.407_{-0.003}^{+0.003}$&$1.08(94)$ \\\hline
116&4639010195&59787.18&570&47.8 &$0.54\pm0.008$&$10.9_{-1.4}^{+1.6}$&$3.18_{-0.43}^{+0.62}$&$0.80_{-0.07}^{+0.07}$&$0.10_{-0.01}^{+0.01}$&$103.6_{-1.2}^{+1.2}$&$0.029_{-0.001}^{+0.001}$&$0.170_{-0.002}^{+0.002}$&$0.289_{-0.002}^{+0.002}$&$0.88(92)$ \\\hline
117&4639010196&59791.24&213&43.6 &$0.48\pm0.009$&$2.9_{-1.5}^{+3.5}$&$24.62_{-12.47}^{+62.09}$&$0.90_{-0.14}^{}$&$0.06_{-0.01}^{+0.01}$&$244.4_{-5.9}^{+5.9}$&$0.019_{-0.001}^{+0.001}$&$0.268_{-0.004}^{+0.004}$&$0.377_{-0.004}^{+0.004}$&$1.14(72)$ \\\hline
\end{longtable}
\endgroup
\begin{list}{}{}
\item[a]{: The model parameters: the optical depth $\tau$,  the  electron temperature $kT_{\rm e}$,  the cover factor $f_{\rm sc}$,  the accretion disk  temperature   $kT_{\rm disk}$ and  the inner disk radius   $R_{\rm diskb}$ at a distance of 7.56 kpc and in the face-on scenario (inclination angel $\theta$=0), the bolometric flux of the diskbb $F_{\rm diskbb}$, the bolometric flux of the corona $F_{\rm diskbb}$, and the total  bolometric flux  $F_{\rm total}$ are in units of $10^{-9}~{\rm erg/cm}^{2}/{\rm s}$, reduced $\chi_{\nu}^{2}$ and the degree of freedom (d.o.f.).}
\item[b]{: Please note that the 'propeller effect' time is between {\bf obsids 4639010184 and 4639010185.}}
\end{list}

\clearpage

\end{landscape}
\clearpage
  


\clearpage

\end{document}